\documentclass[thesis=M,english]{FITthesis}[2012/10/20]

\usepackage[utf8]{inputenc} 

\usepackage{graphicx} 

\usepackage{dirtree} 
\usepackage{wrapfig}
\usepackage{listings}
\usepackage{xcolor}
\department{Department of Software Engineering}
\title{Cross-matching Engine for Incremental Photometric Sky Survey}
\authorGN{Jiří} 
\authorFN{Nádvorník} 
\author{Jiří Nádvorník} 
\authorWithDegrees{Bc. Jiří Nádvorník} 
\supervisor{RNDr. Petr Škoda CSc.}
\acknowledgements{Most of all I would like to thank my supervisor Petr Škoda for patience and willingness to help me with astronomical aspects of this thesis. I would also like to thank Filip Hroch for his support in Munipack usage as well as in theoretical aspects of my thesis. The next person that supported it very helpfully is Martin Reinecke from the HEALPix support and last but not least I would like to thank Markus Demleitner from GAVO for his support far exceeding the scope of GAVO DaCHS package we are using. We also acknowledge grant GACR 13-08195S.}
\abstractEN{For light curve generation, a preplanned photometry survey is needed nowadays, where all of the exposure coordinates have to be given and don't change during the survey. This thesis shows it is not required and we can data-mine these light curves from astronomical data that was never meant for this purpose. With this approach, we can recycle all of the photometric surveys in the world and generate light curves of observed objects for them.

This thesis is addressing mostly the catalog generation process, which is needed for creating the light curves. In practice, it focuses on one of the most important problems in astroinformatics. This is clustering data volumes on Big Data scale where most of the traditional techniques stagger and we have to search for new paths to achieve our goal. We consider a wide variety of possible solutions from the view of performance, scalability, distributability, etc. We define criteria for time and memory complexity which we evaluated for all of the tested solutions. Furthermore, we create quality standards which we also take into account when evaluating the results.

We are using relational databases as a starting point of our implementation and compare them with the newest technologies potentially usable for solving our problem. These can be noSQL Array databases or transferring the heavy computations of clustering towers supercomputers by using parallelism.}
\abstractCS{Pro získávání světelných křivek je dnes potřeba předem naplánovat přehlídku oblohy, kde jsou pevně dané souřadnice expozic, které se v průběhu přehlídky nemění. Tato práce ukazuje, že to není nutné a že lze vytěžit světelné křivky z astronomických dat, která k tomu vůbec nebyla původně určena. Tímto způsobem můžeme zrecyklovat všechny fotometrické přehlídky na světě a vytvořit k nim světelné křivky pozorovaných objektů.

Tato práce se zabývá zejména způsobem generování katalogu objektů, který je nutný pro výše zmíněné světelné křivky. V praxi se soustředí na jeden z největších problémů v astroinformatice. Jedná o klastrování datových objemů na úrovni Big Data, kde většina tradičních technik selhává a my musíme hledat nové cesty k dosažení cíle. Zabýváme se širokou škálou možných řešení z pohledu výkonu, škálovatelnosti, distribuovatelnosti, atd. Definujeme kritéria pro časovou a paměťovou složitost, která jsme vyhodnotili u všech testovaných řešení. Dále si vytvoříme kvalitativní nároky, které také zohledňujeme v hodnocení výsledků.

Používáme relační databáze jako výchozí bod implementace a srovnáváme je s nejnovějšími technologiemi potenciálně použitelnými pro řešení problému. To mohou být noSQL Array databáze nebo přesunutí výpočetně náročných fází na superpočítače s použitím paralelismu.
}
\placeForDeclarationOfAuthenticity{Prague}
\keywordsCS{astronomie, paralelismus, klastrování, big data, databáze}
\keywordsEN{astronomy, parallelism, clustering, big data, database}
\declarationOfAuthenticityOption{4} 

\begin{document}



\setsecnumdepth{all}
\chapter{Introduction}
In this chapter we will define the topic of this thesis, it's contents, justification and purpose. As already said in the abstract, the main part of this thesis is to take a set of observations\footnote{By the term observation, we mean a light dot identified on an image of the sky} and cluster them based on Euclidean and other metrics.

\begin{figure}[h!]
  \centering
    \includegraphics[width=1\textwidth]{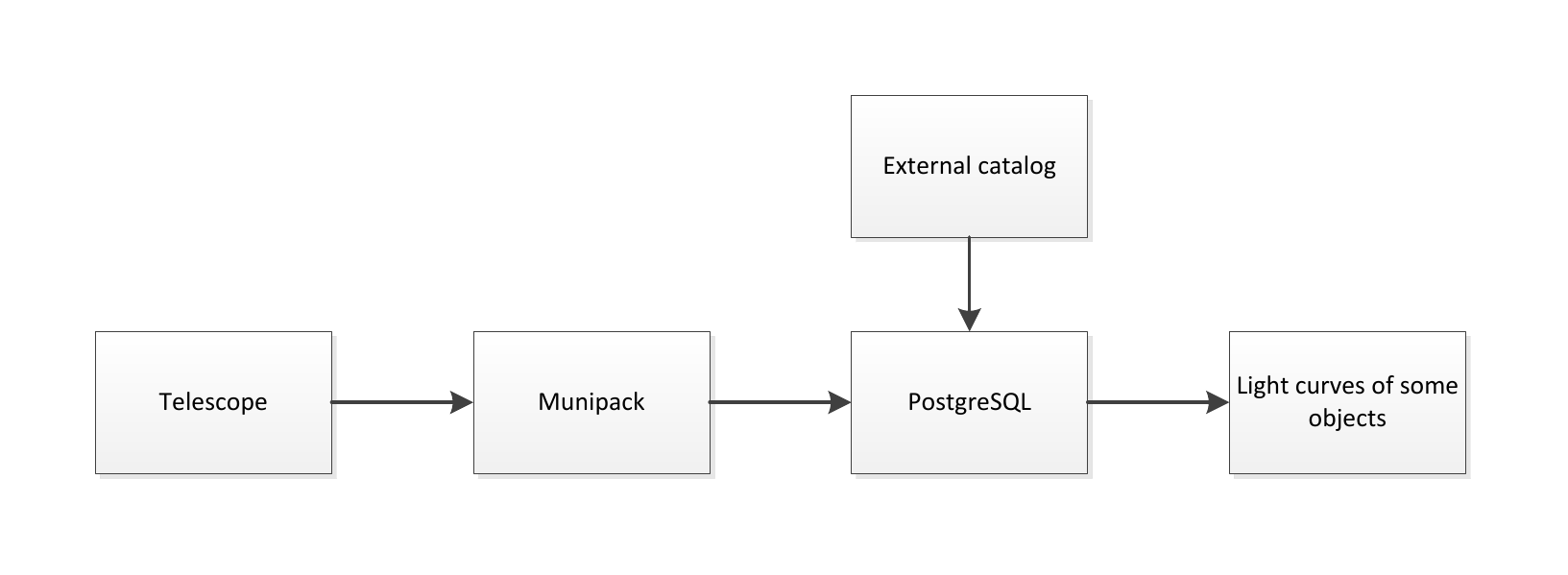}
 \caption{As-Is state}
  \label{figureWFOld}
\end{figure}

\begin{figure}[h!]
  \centering
    \includegraphics[width=1\textwidth]{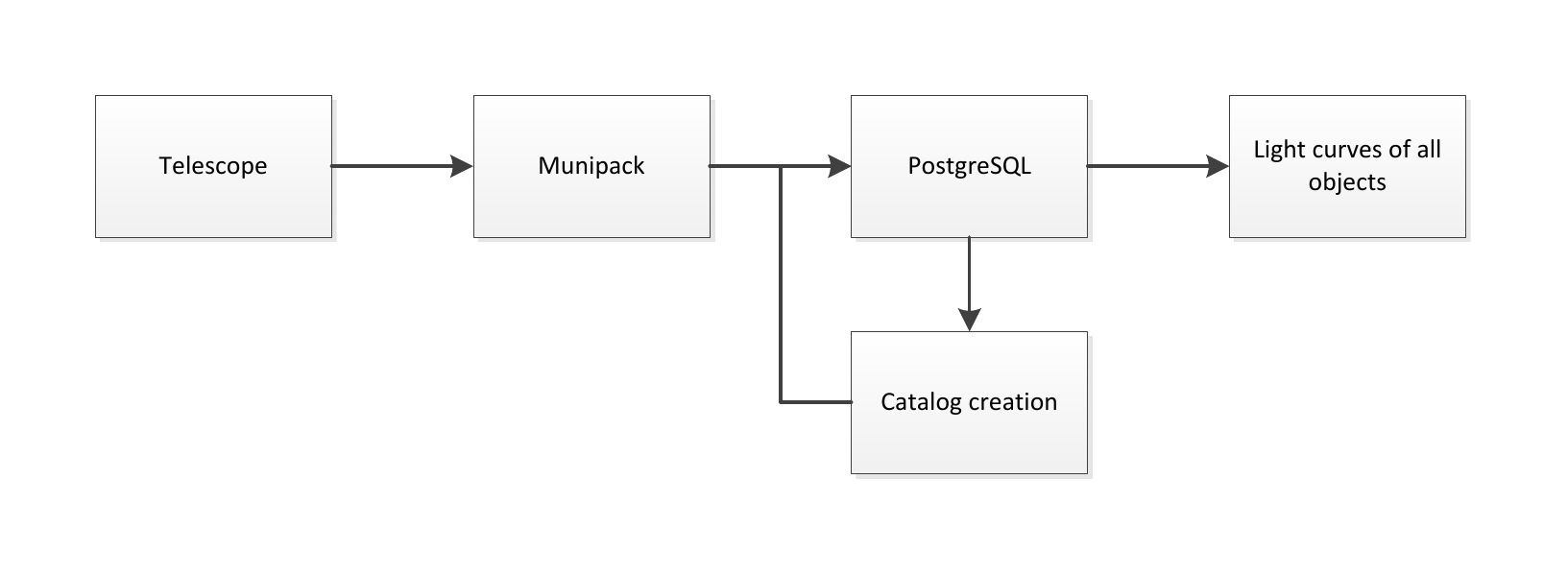}
  \caption{To-Be state}
  \label{figureWFNew}
\end{figure}

The ultimate goal is to provide light curves\footnote{Light curve is a graph of an observation's brightness based on time when the observations where taken.} of astronomical objects to the end user, with as much quality as possible. For that, we need to generate our own catalog\footnote{Catalog of astronomical objects is list of celestial objects identified by their coordinates. These can be planets, stars, galaxies, quasars, etc.} of objects.  The main problem with cross-matching\footnote{Cross-matching in astronomy is understood as a process of matching one data set with another. The criterium is mostly distance between the data points, which means a point from set A will only be matched to a point from set B if their distance is smaller than the criterium.} matching only an existing catalog is that we won't be able to create light curves for all of our observations. If we create our own catalog, however, we can guarantee, that all of our observations will be part of a light curve. The As-Is state can be seen on Fig.~\ref{figureWFOld} and To-Be on Fig.~\ref{figureWFNew}.

\section{Motivation}
In this chapter we will justify the motivation for our thesis. We will explain the usefulness of the work and the benefit it will bring to the end users. The end users for our system are astronomers, but the nature of our data makes it very interesting for geospatial science too.

The high level motivation is to create a light curve catalog for OSPS (Ondřejov Southern Sky Photometry Survey) project~\cite{osps}, with the data originating form the Danish 1.54-m Telescope~\cite{dk154}.  This photometry survey contains hundreds of thousands of images, but almost all of them are observing the same region on the sky.  Identifying celestial objects on these images is a complicated process of astrometry\footnote{Astrometry is a process of measuring exact positions and movements of celestial objects} and photometry\footnote{Photometry is a process of measuring the flux, or intensity of an astronomical objects electromagnetic radiation}. The output of this process is a set of observations of all identified objects\footnote{An observation of individual object is defined by their sky coordinates, photon flux, and lots of other parameters} for each image. 

Our data used for producing the light curves is not actually meant for that originally and that's why we differ from the standard solutions nowadays. This aspect will be brought in detail in~\ref{secDataStructure}.

\section {Data structure}
\label{secDataStructure}
There is a lot of sky surveys which clearly had to solve this problem already. To answer this question decisively, we need to introduce the structure of our data.

\begin{figure}[h!]
  \centering
    \includegraphics[width=0.8\textwidth]{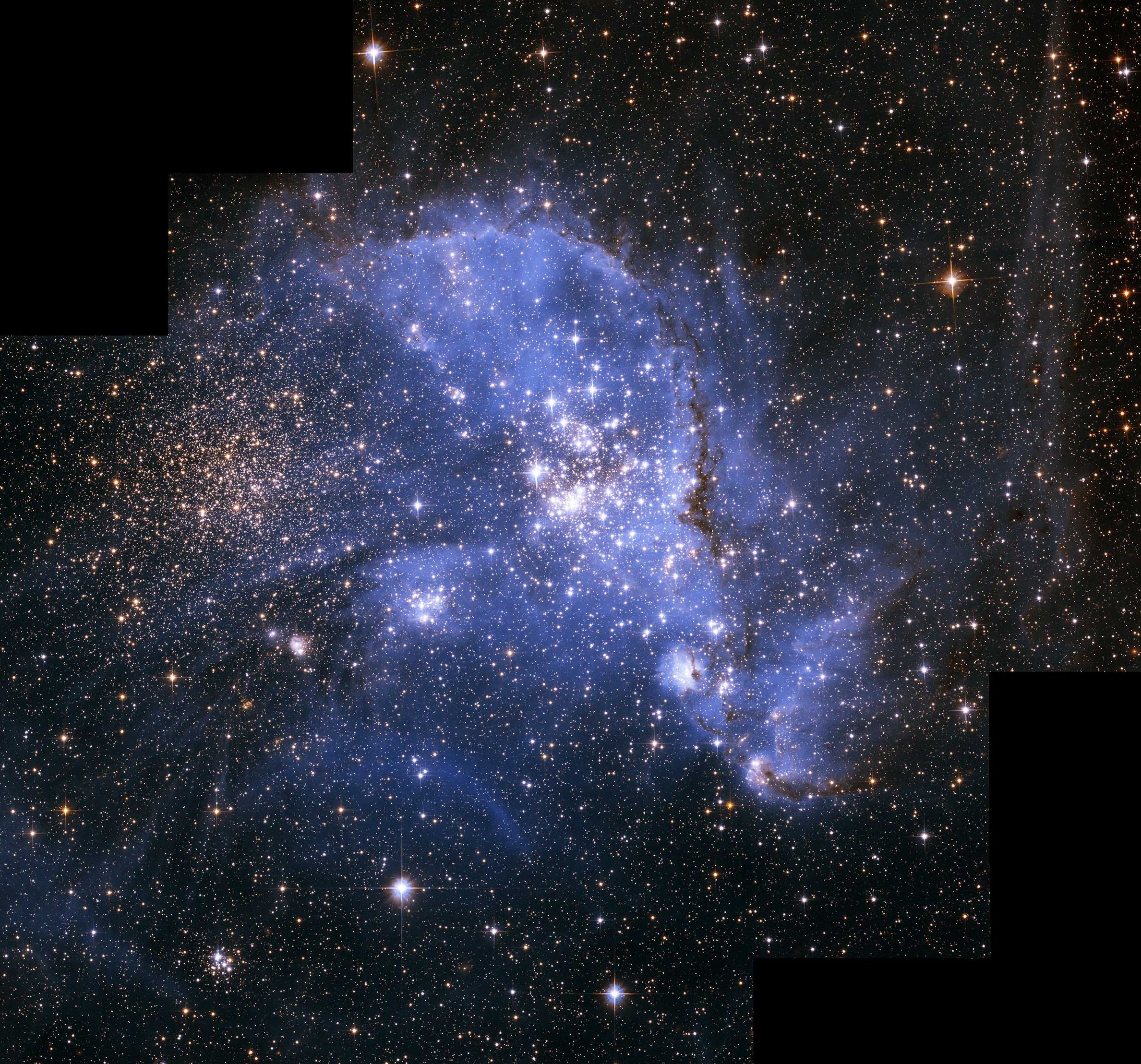}
  \caption{Small Magellanic Cloud}
  \label{figureSMC}
\end{figure}

\begin{figure}[h!]
  \centering
    \includegraphics[width=0.8\textwidth]{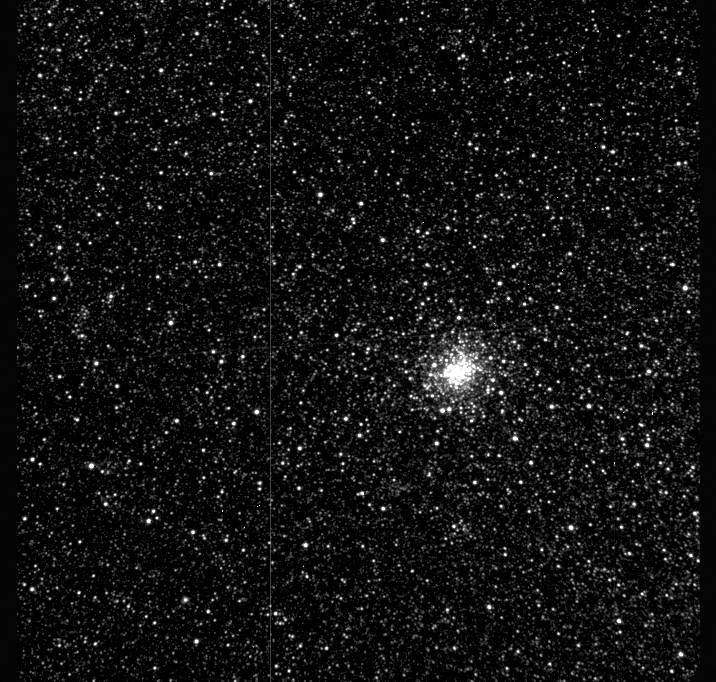}
  \caption{NGC330 image}
  \label{figureNGC330}
\end{figure}

As most of the fields are located in the area of Small Magellanic cloud\footnote{Small Magellanic Cloud is a dwarf iregular galaxy, one of the closest neighbors to our Milky Way} seen on Fig.~\ref{figureSMC}, they are very dense. On one image we have cca 5000 - 25000 identified observations.  For comparison, on an average region on the sky with the same coverage and deepness, there will be between 2000 and 5000 observations. We have cca one hundred thousand images in our dataset, with a total of four hundred million identified observations. The dataset will still grow until the end of the project when we expect to have six hundred million identified observations which we need to assign to celestial objects. A typical image looks like NGC330\footnote{NGC330 is a open star cluster in Small Magellanic Cloud} on Fig.~\ref{figureNGC330}. 

\subsection{Typical versus our data}
In this section we will compare the typical approach to creating a light curve catalog with the one we had to choose for ourselves. The reason for it is the different structure of typical and our data.

\subsubsection{Typical data}
A typical light curve survey will have a before-hand defined grid of image areas which will be observed repeatedly in regular time intervals. This grid will not change during the project. This means that two images taken at different times which cover the same region will overlap almost entirely (the intersection of their coverages will be almost as big as the whole image). This means that a differential photometry and astrometry can be used. The most crucial condition for a differential astrometry to to be successful when comparing two images is to have enough common coverage, so we can match their positions. The difference between absolute and differential astrometry is shown at the beginning of this presentation~\cite{diffAstrometry}.

\subsubsection{Our data}
\begin{figure}[h!]
  \label{figCoverage}
  \centering
    \includegraphics[width=0.8\textwidth]{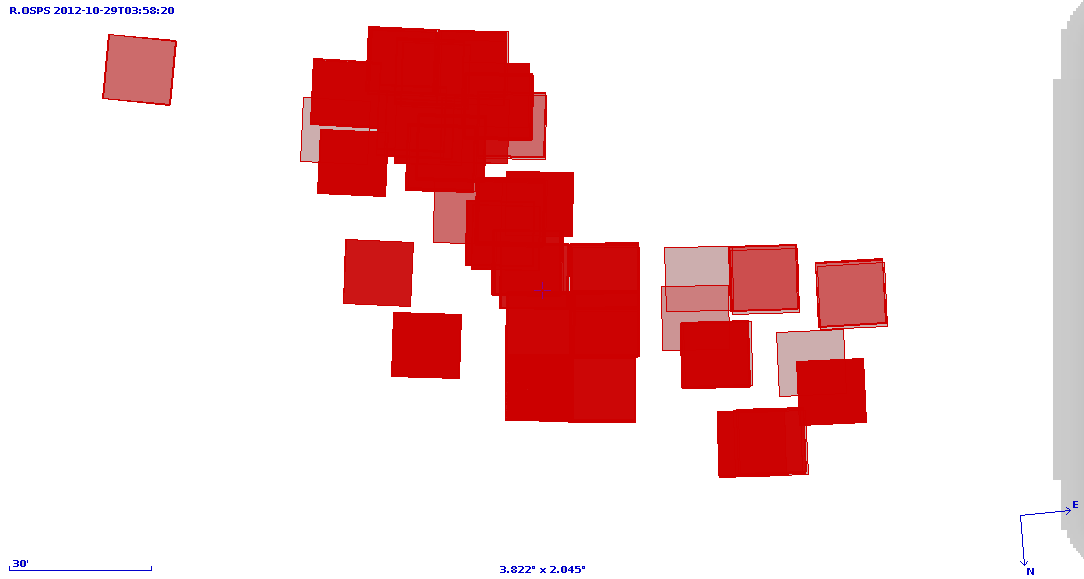}
  \caption{OSPS coverage}
\end{figure}

Unfortunately, we cannot use differential astrometry for our survey. The images taken in OSPS have mostly only few objects of interest (a planet, asteroid, or a Be star, ...), which means two things. First, 99 \% of the data is not originally meant to be used and second, there is no grid for restricting the image positions. This is best seen on Fig.~\ref{figCoverage} -  here we can see the chaotic spread of our images which makes differential astrometry and photometry stagger, because the images won't have enough referential stars in common.

\section {Our solution}
\label{secOurSolution}
We would like to use this 99\% of our data which is just thrown out, but has no less quality and can easily lead to new discoveries. We believe that our survey is not a rare case when most of the data is unused. With our approach, we can actually recycle all of the images in the world even if they were not originally meant for producing light curves and data-mine much information from them.

Another reason why we'd like another approach is that we want our survey to be incremental. That is not always the case for standard surveys, as when they are closed, it would be very complicated to add later (or sooner) taken images to the survey. When new images are taken, the light curves are just updated, not generated anew.

Our astrometry is quite similar to the differential one, but we are not comparing our images directly. It is done with the help of package Munipack ~\cite{munipack}. Instead of computing the astrometry and photometry for all images at once, we create our own set of calibration stars selected from an on-line catalog (currenty UCAC4~\cite{ucac4})  and try to calibrate our image's coordinates with these. We do the astrometry separately for each image, so at this point we don't mind whether the images actually overlap with each other or not.

\begin{figure}[h!]
  \centering
    \includegraphics[width=0.8\textwidth]{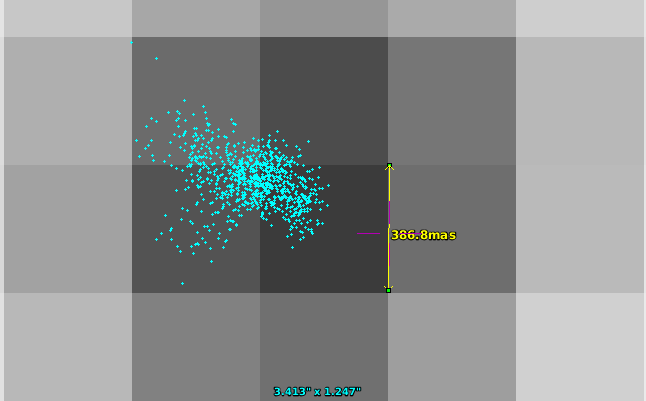}
  \caption{Astrometry accuracy}
  \label{figAstrometryAccuracy}
\end{figure}

This creates a small random error, which will cause all observations of one immovable object create a group of points with a Gaussian distribution of coordinates. The accuracy of our astrometry is around 0.2 - 0.3 arcsec, which is about half the size of our pixels as can be seen on Fig.~\ref{figAstrometryAccuracy}. These groups of points are our clusters, and assigning the cluster identifiers to these points observed on the sky lets us to query for a light curve of one individual object on the sky.

The product of the astrometry is just a binary table of observations for each image. When we take this data from all of the images, we have four hundred million observations which we need to cluster and assign to real physical objects (cluster them). And how we do that is the topic of this thesis. 

\subsection {Incremental survey and Transients}
With our research, we can even create incremental survey. On the first iteration, we cluster all of the observations we have and create the first version of our catalog. The individual star observations on each image are assigned to objects in this catalog during the process.

After our survey has expanded, we take the images that were not processed yet and try to cross-match them to our previously created catalog. For the ones that did not match, we just run the catalog generation process separately, update the catalog with new identifiers, and assign the individual star observations.

This way we can even detect transients, such as supernovas, for which we didn't have a catalog identifier before. This is a great advantage against usual approach, when we just try cross-match against an on-line catalog and throw away the observations of objects, which are not in the catalog.

\subsection{Publication}
The results of this project will be presented on the IVOA Interoperability Workshop – Spring 2015 meeting ~\cite{interop}.

\chapter{Review of possible solutions, prove of concepts}
As there exist no solutions which are solving exactly our problem, this chapter will not be a typical analysis. Instead we have to at least partially implement each possible solution as a prove of concept and then evaluate the result and decid whether this particular solution is feasible. 

At the beginning of this chapter we define the background environment where we have to apply our solutions. There will be also all of the solutions which failed to some aspects. These solutions are analyzed carefully so we don't throw away an already partially or fully implemented solution. 

The background needed will be defined in sections~\ref{secGeneralBack} and~\ref{secDBBack}, all inacceptable solutions in sections Pure SQL~\ref{secPureSQL}, Array Databases~\ref{secArrayDB}, Apache Spark~\ref{secApacheSpark} and the final accepted solution is at the end of this chapter in section~\ref{secCPP}.

\section{General background}
\label{secGeneralBack}
There are many possibilities of how to store astronomical data and publish it to the world. The infrastructure we are using is inherited from my Bachelor's thesis~\cite{bakule}. The general data flow can be seen on Fig.~\ref{figureWFNew}. A more detailed view can be seen further on Fig.~\ref{figPipeline}.

\begin{figure}[h!]
  \centering
    \includegraphics[width=0.9\textwidth]{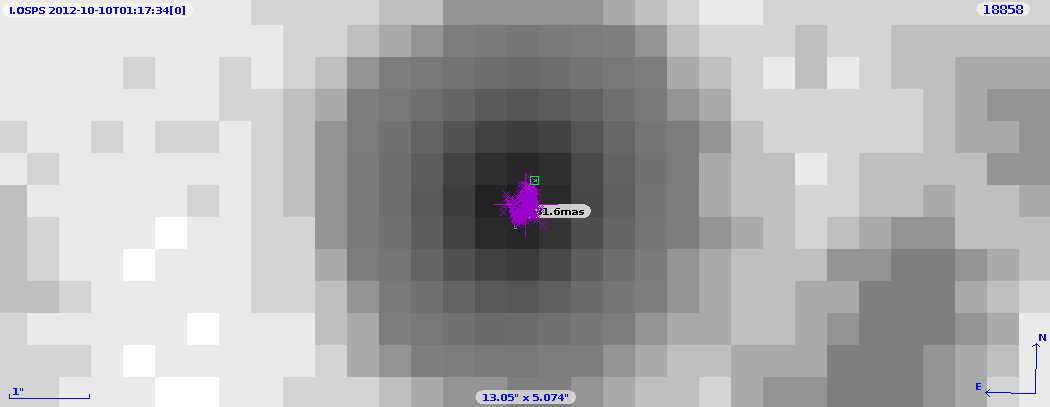}
\caption{Typical star observation}
  \label{figStarCluster}
\end{figure}

\begin{figure}[!h]
  \centering
    \includegraphics[width=0.8\textwidth]{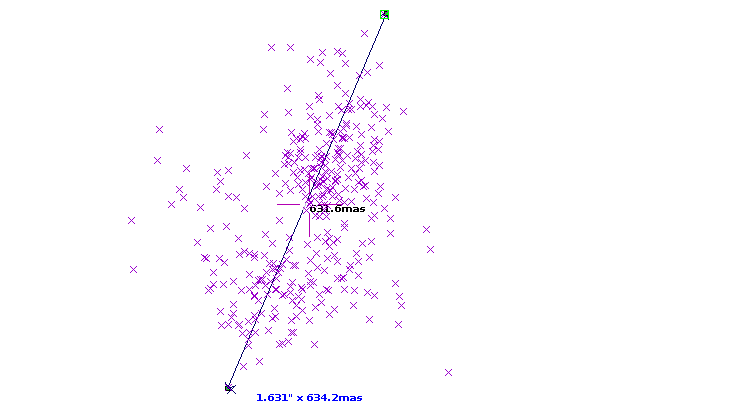}
\caption{Cluster of star observations}
  \label{figClusterClose}
\end{figure}

The main thing we will be focusing on in our thesis is how to properly cluster the astronomical data to produce desired light curves.  We can see a star on Fig.~\ref{figStarCluster} with cca 300 observations distributed around the star center. There is a closer look on Fig.~\ref{figClusterClose}, where we can see observations of one object over a period of time. The error in the astrometry here is around 0.3 arcsec, forming a cluster with a diameter cca 0.5 arcsec, as pointed out with the gauging line.

\section{Database background}
\label{secDBBack}
\begin{figure}[h!]
  \centering
    \includegraphics[width=1\textwidth]{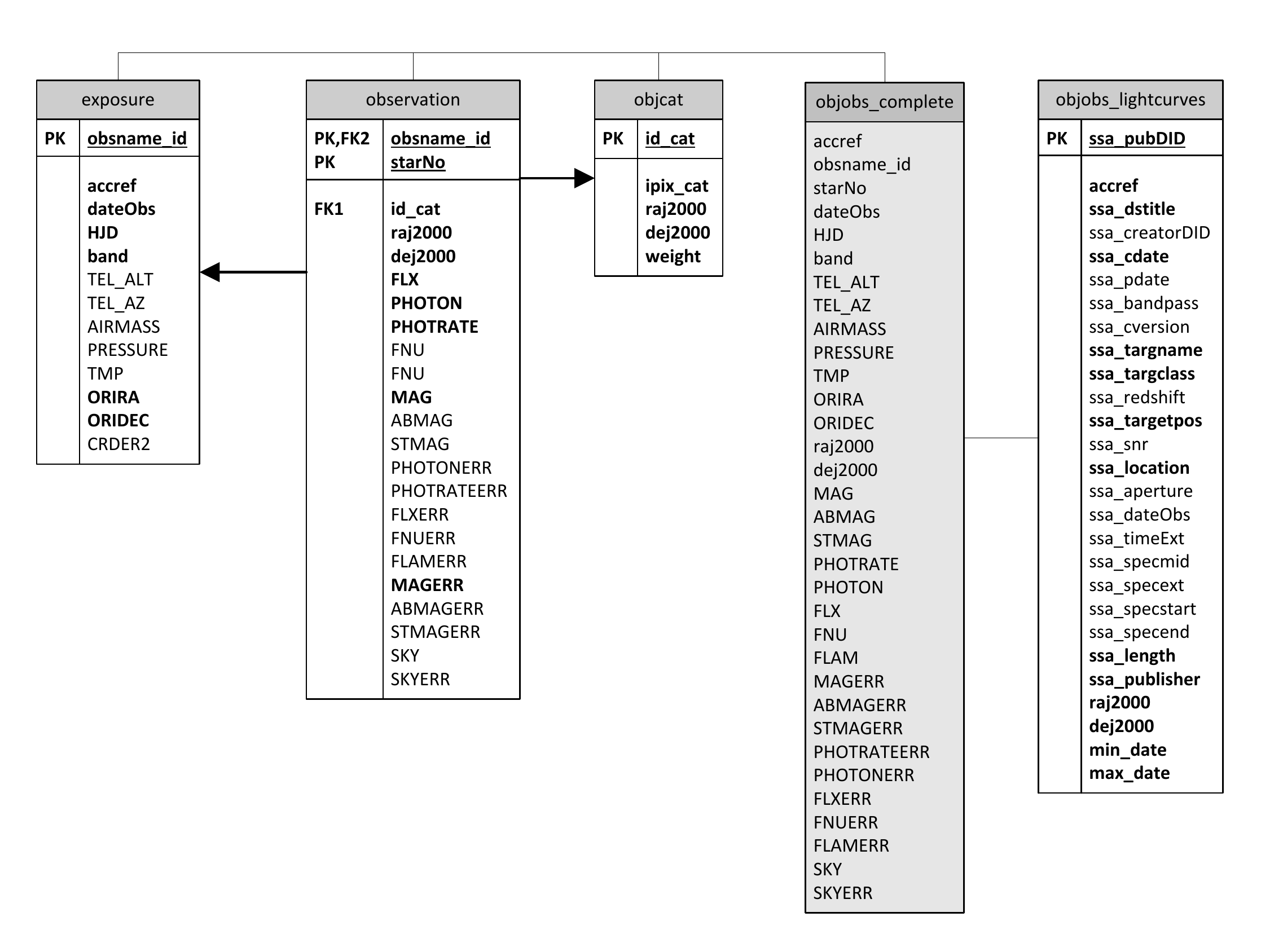}
  \caption{DB Model}
  \label{figDBModel}
\end{figure}

The database model of the underlying architecture can be seen on Fig.~\ref{figDBModel}. Each image is represented by 1 row in exposure table. Observations identified on this image will be kept in observation table, tracked by \emph{obsname\_id} to the exposure they were taken with. Then we will have to create identifiers for the actual objects and write them to \emph{objcat} table. Each observation should be assigned to a catalog object via \emph{id\_cat} foreign key. 

The \emph{objobs\_complete} view is used for easy access to the complete information about an observation. It joins data from exposure, observation and objcat tables, effectively linking image data (when the observation was taken, with which filter) with the actual observation (point on the image with all it's identified attributes) and the catalog identifier (to which real object this observation corresponds).

The \emph{objobs\_lightcurves} table based on the \emph{objobs\_complete} table and is used with the SSA protocol~\cite{ssap} to publish the light curve. This process is described in detail in my Bachelor's thesis~\cite{bakule}.

\section{Pure SQL solution}
\label{secPureSQL}
For summing it up, we are using PostgreSQL database for storing all the observation data as it is nicely supported and used in the astronomy area, has multiple sphere indexing algorithms implemented and can easily handle the amounts of data we are using. The main reason, however, why we are using this architecture is the GAVO DaCHS ~\cite{gavoDachs} package, which we are using for the data ingestion and publishing, and this package is built on the PostgreSQL database. 

In this section we will describe all of the solutions which try to process all of the work inside the PostgreSQL database.

\subsection{PPMXL Catalog}
\begin{figure}[h!]
  \centering
    \includegraphics[width=0.8\textwidth]{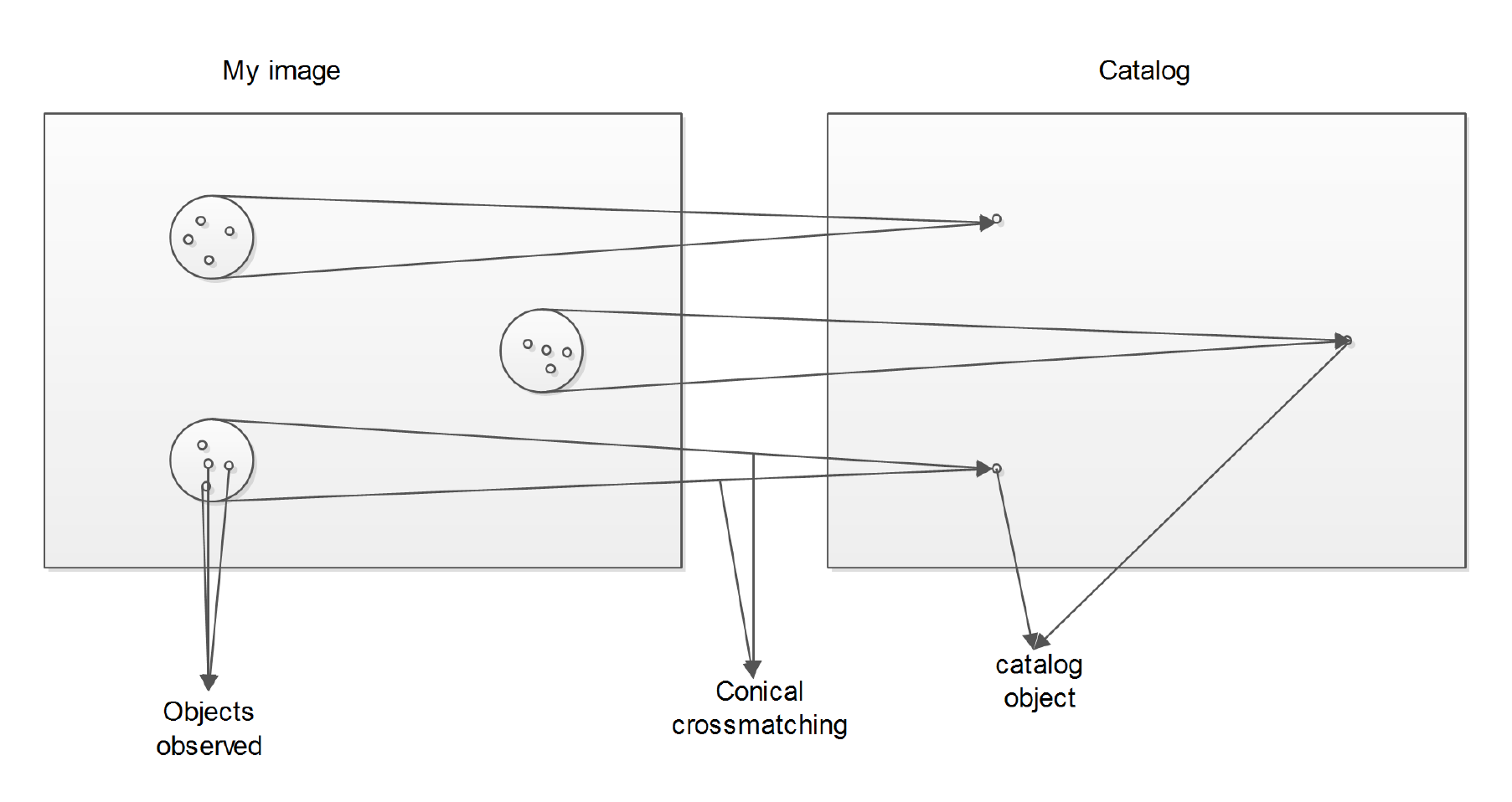}
  \caption{Catalog cross-match}
  \label{figCrossMatch}
\end{figure}

In the original solution described in my Bachelor's thesis~\cite{bakule}, we did not create our own identifiers. Instead, we took our observations and tried to cross-match them with a deep enough on-line catalog. The idea is illustrated on image~\ref{figCrossMatch}. The individual observations on the left (one cluster is the same one as on Fig.~\ref{figClusterClose}) are cross-matched to catalog objects on the right.If an observation has no matching object in the catalog, it has no other way of assigning itself to a light curve and will be forgotten.

The best results were produced with the help of PPMXL catalog~\cite{ppmxl}. We managed to identify cca 70 \% of our observations and within the 70 \% there were still errors mostly caused by duplicate entries in the catalog. Throwing away more than 30\% of our data is alone an unacceptable drawback.

\subsection{"Silver bullet" query}
\label{secSilverBullet}
So we need to create our own catalog. The possible "integration friendly" solution is to write a simple SQL query that would extract the identifiers we need. Because we already have the data ingested into a relational DB (PostgreSQL), it would be very convenient if we could create these identifiers inside the database, without the need of moving big amounts of data to external applications. 

Thus, we created a \emph{"silver bullet"} query, which is just taking the data from \emph{observation} table, grouping them on some criteria, and writing the catalog identifiers to \emph{objcat} table and the assignments of points via \emph{cat\_it} field back to the \emph{observation} table. 

\subsubsection{One big query}
The query is based on the fact, that our clusters have cca 1 arcsec diameter. So we are grouping the data on a condition, that all of the points in one group can be connected by distances smaller than 1 arcsec. In practice this is implemented by a self-join which iterates over all points and for each one of them creates a group of points which are closer than 1 arcsec. Then if we iterate over groups of these points (members of my group), we are just observing this cluster from different points of view. From these we can choose the one that has the "best" view - meaning he sees the most points - has the largest group. This means just that the point is the closest to the center of the cluster and if we create an average of coordinates of all his neighbors, it will lie precisely in the center of this cluster. This approach can solve the corner cases too (e.g. three points in one line or a equilateral triangle).

This clustering algorithm can be implemented in PL/SQL, runs fast and has a very high precision for most of the patterns the observations  can create on the image. 

If we look at the complexity of this algorithm, we see that it is highly dependent on number of points in one cluster. If we are unlucky and we have a cluster of 1000 points which have exactly the same coordinates, for each one of them we get 999 coordinates with a distance zero. We have an \(O(n^2)\) complexity where n is the number of points in one cluster. This counts for both computational and memory complexity.

Aside from the fact that our data has high density (on one image), it has also high density in the sense that we have a lot of overlapping images. We have at most cca 1000 points per cluster, at average cca 100. On the example of the whole dataset which has cca four hundred million points, one most basic point represented by 16B (right ascension, declination double precision), that means cca \(6 GB * 100^2 =  600  TB\) of intermediate results. As we would like our algorithm to work for even bigger data sets, this is a major drawback of \emph{"silver bullet"} query algorithm.

\subsubsection{Parallel smaller queries}
Because the complexity problem is fatal only for the memory, parallelizing the query for smaller chunks of data will actually solve this problem. However, that creates another problem - how to divide the data into chunks? 

The Q3C~\cite{q3c} IPIXes\footnote{IPIX is a Q3C identifier for one point on the sky. It is a long integer.}, which we are using for our observations, can be used for sorting the data and then just slicing the chunks sequentially. The spatial locality of these identifiers is quite good, but is not guaranteed (close points on the sky will have their IDs usually close to each other, but not always). We also have high probability of slicing the clusters on the edges of the chunks, and together with not guaranteed locality for the IDs, we will produce duplicate clusters close to each other. This problem can be reduced, if we iterate over points from the chunks only and look for their neighbors in the whole \emph{observation} table.

If we were to accept these duplicate errors and continue with the testing, we will encounter the final bottleneck of this solution. The smaller the chunks are, the less memory we will use (and the more threads we can actually use on one machine). But as we are already pushing PostgreSQL to the limits with the actual implementation (PL/SQL functions, looping over the table manually), the planner is very confused and will not produce any reasonable query plan (which would be loading the actually processed chunk and the whole observation table index into memory). Instead, it will do a lot of random disk seeks when searching the index for close neighbors and then using the whole memory for storing intermediate results. This query realization cannot be changed easily without changing the PostgreSQL source code. And distributing a PostgreSQL database over several disk nodes is very complicated and with already limited query plan quality, would not probably solve the problem.

\subsubsection{"Silver bullet" query summary}
The \emph{"silver bullet"} query solution is working really well with small data, but we encounter very big problems with scalability. This solution is limited either by disk capacity (storing intermediate results) or by the disk speed of random seeks (when parallelizing the query into smaller parts) and these limitations cannot be overcome, so we need to search for other solutions.

\subsection{Iterative query}
\label{secIncrementalQuery}
The limitation of \emph{"Silver bullet"} query is not a problem of the SQL, it is just too high complexity of the actual clustering algorithm used. We can create a streaming algorithm which can process the data with a linear complexity. We will call this approach an \emph{Iterative query}.

The idea is a quite naive sort of K-means algorithm. We iterate over all of the points in the database and for each one process the following condition.

Do I have a catalog identifier in a given range within my coordinates? If no, then I am a new cluster and the identifier is me. If yes, I add myself to that cluster and just update the cluster coordinates by a weighted average of mine coordinates and the ones already in that cluster.

This approach does not solve the corner cases when we don't iterate over the points in the right order. Example pattern here can be a cluster of directly 500 mas diameter. We start with a point completely on the left and create new cluster. Then it happens we take the second point completely on the right. The distance to the previously created catalog ID is exactly 1 arcsec far away, so we have to create a new cluster. Then as we iterate over the rest of the points, some of them will be assigned to the left cluster, some to the right. In the end we have 2 clusters close to each other instead of one whole.

These cases are in practice quite rare though, so we can accept them as error rate of this implementation. There is, however, a bigger drawback here which comes from using SQL. This language is just not built for looping over each single row of the data and doing operations on such level. For each point we have to do a couple of random seeks in the spherical database indexes before we find all of the points closer than the distance limit. With an average of couple ms per one cycle (which is really fast for close neighbors lookup in a relational database), it takes us weeks to process our whole dataset with very high disk usage the whole time. Under such load, we even discovered that PostgreSQL is quite unstable, because it was simply not meant for such usage.

\subsubsection{Iterative query summary}
Even if all other arguments were beneficial, we simply cannot accept instable solution. Another option is of quite different matter.

\subsection{IPIX iterating}
One quite different approach still operating on the database level is the following. Instead of iterating over the points we can iterate over some measure defining the clusters itself. If we create a grid of squares with the same resolution as the cluster size, we can actually iterate over the grid and just ask what points are in this column. A sky indexing plugin in called Q3C~\cite{q3c} can actually do that quite efficiently. We call the strategy \emph{IPIX iterating}. 

\begin{figure}[h!]
  \centering
    \includegraphics[width=0.8\textwidth]{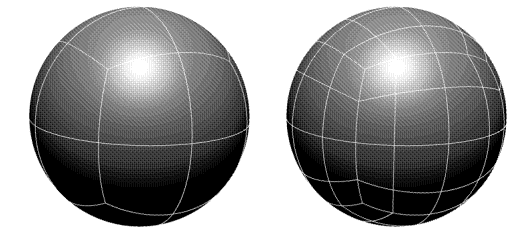}
  \caption{Q3C Pixels}
  \label{figQ3C}
\end{figure}

On this Fig.~\ref{figQ3C}, we can see how the sphere is partitioned to "squares" by the Q3C algorithm. It is based on quad tree cube partitioning of the sphere as can be seen on image~\ref{figQ3C}. 

After assigning points to the squares in the grid, we can join the ones which have points assigned and are next to each other, as that probably means that we sliced a cluster which lies on the borders of these squares. This approach however costs us precision in higher density areas where the cluster distances are comparable with the cluster sizes as we join more clusters together under one catalog identifier.

The big plus for the IPIX iterating method is it's speed. For the whole dataset (cca four hundred million observations) , it runs cca 40 minutes. It's complexity is \(O(n)\) where n is the number of Q3C squares used for the sphere partitioning.

The accuracy is about the same as if we used the on-line catalog - around 70\% of the object identifiers are identified correctly.

\subsection{Combination}
It would be very nice, if we could take advantage of such a fast algorithm, which provides high precision for cca 70\% of our observations and falls off only where the star fields take up in density (clusters closer to each other). We could separate the rest of 30\% and process them in some other slower algorithm. The problem is, that 30\% will still take days to process with the iterative query~\ref{secIncrementalQuery} link and with the \emph{"silver bullet"} query described in chapter~\ref{secSilverBullet} we wouldn't help ourselves, as it already solves sparse fields easily and has problems with the dense ones.

The biggest problem is that telling whether we actually joined the intermediate results correctly is actually as hard as the clustering problem itself. We tried several heuristics to separate the erroneous results, but every time with worse results than with other non-combination approaches. So combining the \emph{IPIX iteration} with other sorts of clustering algorithms is a nice idea, but we didn't manage to solve to create a simple fast metric which would select all of the incorrectly assigned points or incorrectly identified clusters.

\section{Array Databases}
\label{secArrayDB}
That's it, we are done with the relational databases. Clearly they are not meant for such kind of work and the argument of not transferring big amounts of data seems not that important after all. With freeing ourselves from the bonds of standard database, we can look for more exotic ones. 

The array databases have very nice way of storing data. In general we are talking about column store (instead of standard row store) of data. The data is logically stored as sparse matrices, where we can search for close neighbors very effectively. That suits our data really well, as we have at least two dimensions (right ascension, declination coordinates) where we would really benefit from this strength. The array databases are even more effective, if we have more dimensions of the data (so we can add more parameters for our clustering). Another strength of array databases is that the sparse matrices can be easily sliced and partitioned between multiple physical database nodes. 

The biggest advantage is that most of the solutions are highly tuned to keep most of the data in memory rather then disk, so we would have no more problems with disk random seeks when searching for closest neighbors.

We could even group our data spatially, assign them to concrete partitions and then run the clustering algorithm separately, knowing that all of the points we need are actually in our physical partition. When we add a small overlap to these regions on the sky, we will be sure, that no clusters will be split on the edges. 

With this approach we could even use the high support for extensibility of for example \emph{MonetDB} and write the clustering algorithm as an extension to the database core and call it directly from SQL, leaving the parallelisation management to the database itself. Or we could use \emph{SciDB} with a simple C++ program that would transfer the data from database node, process it, and return the results. The database would again care about the parallelization over database nodes, where only a small geometrical chunk of sky will be stored and processed.

We will now have a look at some most famous array databases used nowadays. The \emph{MonetDB}~\cite{monetDB} and \emph{SciDB}~\cite{sciDB}. We will not make a detailed comparison of these databases in the sense of benchmarking or some detailed analysis. We will compare them only from our point of view, which is usability for storing, publishing and most of all clustering astronomical data.

\subsection{Requirements} 
The main points which we demand from the array databases can be summarized into the following:

\begin{itemize}
\item{Installation} - As we are using Debian for our servers and that is not something to be changed in near future, we need a support for this system.

\item{Storage for astronomical data} - We need to be able to store astronomical data at least as well as in PostgreSQL Q3C~\cite{q3c} spatial indexing schemes.

\item{Publishing of astronomical data} - If we decide to migrate to array databases, we have two possibilities. We might migrate only part of our data and functionality, or we migrate all of the functionality. It would be very complicated to integrate the array database with PostgreSQL in order to keep part of the data here and part of the data there. When we decide to migrate fully though, we need to implement the protocols for publishing data on top of an array database.

We are currently relying ourselves on the GAVO DaCHS package~\cite{gavoDachs} which implements majority of the IVOA VO~\cite{ivoa} protocols and works currently with PostgreSQL only. If we cannot transfer the data model without any significant changes, it will be very difficult to accommodate the processes above them accordingly.

In the end, we decided not to use the array databases only partially, as that would bring us more troubles then benefits. As this decision was also connected with other aspects, we won't terminate our analysis here, but try to compare other aspects needed for our cause too.

\item{Data migration from PostgreSQL} - We need an easy data migration from our data centers which are currently running on PostgreSQL.

\item{Extensibility for clustering} - We need to be able to write an extension to the database or call a program from the database which actually clusters the data. Otherwise we would not benefit from the fact, that database can handle the parallel processing on each data partition (i.e. database node) for us.

\item{Fast neighbor lookup} - This is not needed only for the clustering algorithm, as obviously it can run outside the database as a stand-alone program. But we actually need it for any kind of data publishing to the outside world, as most of the astronomical queries will be spatially based.

\item{Easy data partitioning} - We need a simple way to control the partitioning of our data across the database nodes. To be the distributed clustering algorithm truly effective, we need to ensure high spatial locality on the data. Each partition has to represent an area on the sky. In other words, one cluster is permitted to have it's members stored only at one particular data partition,i.e.,physical database node. We would be relying on that if we decided to process the data in smaller chunks. It would be much easier if we could process each one separately without the need of communication with the others.

\end{itemize}

\subsection {\emph{MonetDB}}
We made the most detailed analysis for \emph{MonetDB}~\cite{monetDB} as it is really close to what we need.The \emph{MonetDB} is written in C, supports mainly Linux systems and is very easy to install on Linux. 

\emph{MonetDB} is quite advanced as it has been developed since 1993. Amongst its biggest strengths belongs high performance vertical fragmentation, automatic and adaptive indices and run-time query optimization. In other words, we don't have to worry for example about indexing the data correctly, \emph{MonetDB} will create the indices itself, statistically based on the queries which will be requesting the data.

\subsubsection{Installation}
If we want to install the database on one server, it is very easy. For a Linux-based system, we can install it similarly to any relational database - as a distributed package.

\subsubsection{Storage for astronomical data}
\emph{MonetDB} is based on a an exotic way for data storage. It stores tables using vertical fragmentation (storing each column as one table), called \emph{Binary Association Table (BAT}. Each table is stored using a key-value mechanism, where the keys are always a dense sorted list. Both the keys and values are stored as memory mapped files, which ensures very high performance with data access times. In the values we can store anything - in case of variable-width types the value is separated into into a reference (offset) and the real value of variable length. 

For the right indexing (which in case of \emph{MonetDB} means just sorting) the data, we would need a sphere partitioning algorithm. We would also need to implement the geometrical queries ourselves, whereas for PostgreSQL we can use already tested and reliable plugins like Q3C~\cite{q3c}.

There are actually several User Defined Functions in \emph{MonetDB} under LSST~\cite{lsst} package, based on HTM spatial indexing~\cite{htm}. For all out publishing of astronomical data, however, they are incomplete and last but not least poorly documented. We would not choose HTM indexing for our data either, because it's complexity climbes with the indexing resolution and it does not guarantee the pixels to be of same size.

There is also a project which tried to use \emph{MonetDB} for SDSS survey SkyServer~\cite{sdssSkyServer}, where they actually implemented spatial queries based on zones. This algorithm is described in article ~\cite{zoneIndex}, but these methods deviate quite significantly from our current solutions and integrating them into our clustering algorithms would be very difficult.

\subsubsection{Publishing of astronomical data}
Here comes the real problem. As we mentioned above, transforming the data may be difficult, but it can be done. But it means we would have to rewrite the implementation of the protocols responsible for publishing of our data.

As I have not found any good enough alternative to GAVO DaCHS~\cite{gavoDachs}, it would mean implementing the support for another database layer to this package directly, which is simply more effort than we would like to invest only for the sake of better clustering results when all other functionalities we already have in the current implementation using PostgreSQL.

\subsubsection{Data migration from PostgreSQL}
\label{secMonetData}
We can migrate the data through two channels. First one is through \emph{pg\_dump} utility, which gives us a database dump of all the tables and data. As not all the data types in \emph{MonetDB} and PostgreSQL match directly, we have to manually repair the dump file according to the \emph{MonetDB} data types. There are some open source tools which do that automatically, but for bad experience with such things, we chose another, safer approach.

Second approach is based on export to simple text (e.g. CSV) files, define the tables in \emph{MonetDB} manually, and just ingest the data from CSV. This approach is safer, as we have complete control over the data types used in \emph{MonetDB} and we solve all the conflicts before-hand, separated from the data.

Each PostgreSQL and \emph{MonetDB} have built-in tools which can export and import CSV files, so there is no problem with this requirement.

\subsubsection{Extensibility for clustering}
Construct named \emph{User Defined Functions (UDFs)} are used in \emph{MonetDB}. These functions are mostly written in pure C and wrapped by \emph{MonetDB} inner assembly-like language called MAL. We can write the functions directly in MAL, but that is not advised, as it is not an easily debuggable language. The MAL instruction to which we can link a C function has to be mapped to an SQL function, which can call the functionality directly from the SQL front end.

This architecture is quite complicated, but it allows us to call the BAT functions directly from the C code. It is very convenient, as now we can just pass points which we want to cluster to the C function, and it can write the results directly to several other tables (e.g. cluster IDs and cluster assignments of the original points). These functions are then stored with the \emph{MonetDB} source code, and using the bootstrapping algorithms, they can be included in the whole build and made part of the \emph{MonetDB} distribution.

Originally, we also thought we could use the very interesting \emph{MonetDB} functionality - the actual integration of R to the database itself. As R contains a lot of clustering functions and modules, it would be very nice if we could cluster the data in the database directly. There is also an example of how to use K-means in \emph{MonetDB} using the R module~\cite{monetDBR}.

But - as the conventional clustering algorithms fall off with both time (high complexity) and quality (usually merging clusters which should not be merged) for Big data, we would have to divide the data into smaller chunks which could be computed in parallel. This would be very nice when used in synergy with the database partitioning. We could run the K-means in parallel for small parts of the data where we could ensure high IO bandwidth if each partition would be saved on a separate hard disk. 

\subsubsection{Fast neighbor lookup}
As we already mentioned in the \emph{Storage for astronomical data} argument, the data is stored in separate tables for each column. If we take the two most important ones for clustering - the coordinates right ascension and declination - we can see that the neighbor lookup will be fast even without explicit indexing. 

As each table is stored as a separated head which contains sorted IDs of the actual values, if the data is sorted geometrically the lookup will be very fast. Ensuring this data locality is quite a hard problem, but there is nowadays a number of sky or sphere indexing algorithms, which we can use.

\subsubsection{Easy data partitioning}
In the official documentation of \emph{MonetDB}~\cite{monetDB} there is no word of partitioning support. There are mentions of possibilities and experiments in the SDSS SkyServer paper~\cite{sdssSkyServer}.

We still can use \emph{MonetDB} for clustering the data using the \emph{Iterative query} approach mentioned in section~\ref{secIncrementalQuery}.

We can still sort the data on disk accordingly to the geometrical locality on sky, as this helps us with finding the nearest neighbors. We decided to use HEALPix~\cite{HEALPix} library to take care of that. It has a very nice future that the pixels partitioning the sphere are of equal size, thus the data will be distributed uniformly in the grid. We will be talking about that in the separate chapter~\ref{secHEALPix}. 

We discovered later that such synergy with \emph{MonetDB} is not actually possible, or at least very complicated. The \emph{MonetDB} is written entirely in C so the bootstrapping of the \emph{User Defined Functions} mentioned in \emph{Extensibility for clustering} point can only work with pure C functions. HEALPix library has APIs for C++ as well as for pure C, however, a key functionality needed for the data partitioning is only available in C++. The thing we need here is to be able to ask for neighbors of a pixel to compute overlaps (see chapter~\ref{secHEALPix}).

We could declare the API functions of our C++ program extern C then, but that would still mean we have to build and distribute our C++ application separately from the \emph{MonetDB} code. This is a very high cost we would have to pay and collides with the point  \emph{Extensibility for clustering} too.

\subsubsection{Conclusion}
We have discussed the individual points we need from the database above and here comes the conclusion. Strong points of \emph{MonetDB} are the Installation, data Migration and the fast neighbor lookup. The storage of astronomical data is fine, but we would have to implement our own plugin for querying the astronomical data. We could base this plugin on HEALPix~\cite{HEALPix} data indexing but we could not distribute the plugin as a part of \emph{MonetDB}, as it cannot be written in pure C.
 
In the end, all comes down to the \emph{Extensibility for clustering} requirement. As much as \emph{MonetDB} is opened and easily extensible by pure C functions, it cannot be extended by C++ code.

Overall, \emph{MonetDB} seems quite mature, but the fact of inextensibility by C++ (which we need for HEALPix, more info in chapter~\ref{secHEALPix}) and the lack of horizontal partitioning make it inappropriate for our case.

\subsection {\emph{SciDB}}
Another option we considered here is the \emph{\emph{SciDB}}~\cite{sciDB}. As we realized quite quickly that it does not fit for our cause, the analysis will only be that much thorough to justify our decision.

The \emph{SciDB} is a partially commercial project where the linear algebra module (which we could make very good use of) is available only in the commercial version. In the scientific open-source version there are all of standard functionalities we need though so we will not dismiss it too quickly.

\subsubsection{Installation}
\emph{SciDB} has a very nice installation automation script, which can install it to multiple database nodes (even on different servers) easily. However, the \emph{SciDB} packaged distribution is very sensitive to the OS version. Installation from packages requires RedHat / CentOS, and Ubuntu to the 14.9 release. If we have another system version (in our case Debian 7), we have to install manually from the sources. As we are considering extending these sources, this is not too big obstacle for us.

\subsubsection{Storage for astronomical data}
\emph{SciDB} has been founded as a part of the LSST~\cite{lsst} project originally. In the flow of time, however, they deviated from this original purpose and formed a stand alone partially commercial project. LSST chose to use massive distributed data center based on MySQL nodes thereafter.

The examples of using \emph{SciDB} for geospatial as well as astronomical data can be found over the internet. The logical storage of data is that we have a multi-dimensional array of cells, which can hold an arbitrary number of attributes. These attributes can be of any (even user-defined) type.

Data transformation will be needed, but no more complicated than when using \emph{MonetDB}. Actually it will be a lot easier, as we can define the cells as coordinates and any attributes assigned to those coordinates can be hold in an array of attributes directly.

\subsubsection{Publishing of astronomical data}
The real issue here again comes with the support for IVOA VO~\cite{ivoa} protocols to publish the data. The problem is that even if we managed to change the data model to the better, we simply cannot afford to invest time connected with adding another database support to the GAVO DaCHS~\cite{gavoDachs} package.

\subsubsection{Data migration from PostgreSQL}
As we already decided to use CSV export from PostgreSQL, \emph{SciDB} supports this future fully. The \emph{pg\_dump} file would require corrections of types, the same as with \emph{MonetDB}. See \emph{Data migration} in previous chapter~\ref{secMonetData}.

\subsubsection{Extensibility for clustering}
The \emph{SciDB} supports multiple types of extensions - user-defined aggregates, user-defined array operators and user defined functions. The last are the ones that interest us most. These are scalar functions which accept variable number of arguments of different data types, and return another data type. 

We would have to define our own types for passing the input data (observations) to the function, as well as for the returned value, which would be the original points grouped into clusters. Using parallelism inside this function shouldn't be a problem and \emph{SciDB} currently supports only C++ for user defined functions, which perfectly suits us, as we need it for our HEALPix~\cite{HEALPix} library.

\subsubsection{Fast neighbor lookup}
\emph{SciDB} storage model ensures, that neighbor lookup will be fast as long as the neighbors will be physically close in the multidimensional matrix. We are talking about the cells in this matrix, which are representing point coordinates in the sky.

As we already mentioned above, to enable this fast lookup by ensuring data locality, we need to be able to sort the data accordingly. Thus, using a sphere indexing algorithm like HEALPix~\cite{HEALPix} is needed and the extensibility of \emph{SciDB} allows us to do that.

\subsubsection{Easy data partitioning}
\emph{SciDB} has a very transparent system of logical chunks, which are dividing the multi-dimensional arrays into groups. These can be easily partitioned over separated media. There can be also defined overlaps between these, so if we can implement a clustering algorithm working on one node, it will not require data transfer from other nodes.

\subsubsection{Conclusion}
The strong points of \emph{SciDB} are the data partitioning, extensibility and fast neighbor lookup points. Other points are acceptable, but very controversial for us is the publishing of the actual data. As we don't like to store part of the data in PostgreSQL and part of it in \emph{SciDB}, we would have to integrate \emph{SciDB} with implementation of IVOA VO~\cite{ivoa} protocols. 

\subsection{Array databases conclusion}
As much as we liked the idea of moving a part or our whole data center to array databases because they work with astronomical data in a more natural way than relational databases, the costs are too high. We are not discontent with the nowadays functionality provided by standard database such as PostgreSQL along with it's sky indexing plugins, the only reason we are trying something different is their inappropriateness for clustering algorithms.

The original thought was that we could migrate only the clustering algorithm to the array database, but that is such an overkill if we compare it to a stand alone C++ program. As we agreed it is not profitable to move all of our functionality to the array databases, they are not useful for solving our problem at all.

\section{Apache Spark}
\label{secApacheSpark}
Next thing we considered is using the Apache Spark~\cite{spark} language, as it can be used very efficiently for parallel cluster computing (here by the term cluster we mean server cluster). We dismissed this language because it's linear algebra module has small support for clustering algorithms (in fact there is only K-means algorithm) and if we are using a lower level language, we would like to do a benchmark for multiple clustering algorithms on the data too.

\section{C++ application}
\label{secCPP}
So the last resort for us are the oldest, but still most efficient tools available. We will implement the clustering algorithm as a stand alone C++ application, to which we will transfer the data. PostgreSQL is very fast with CSV exports/imports, so we stick with that idea and the data will be transfered in this format. We chose C++ explicitelly, because it is fast, relatively easy compared to pure C and HEALPix~\cite{HEALPix} library supports it, which we will bring up in chapter~\ref{secHEALPix}

We will not change the original database model, so the input will be only one table of observations, which can be represented by right ascension and declination, plus some unique identifier tracking the observation back to the image file it was identified on.

The output will be a list of catalog coordinates of our light curves with some catalog IDs and a list of the original observations assignments.

As we don't have to limit us because of the database possibilities anymore, we can implement any clustering algorithm we like. We decided to implement the \emph{Incremental query} defined in~\ref{secIncrementalQuery} as a base for our benchmarks, because if we manage to load all the data set into the memory, there will be no problem with iterating over each particular point.  This algorithm has it's own limits though, and as there are lots of C++ clustering libraries available, we can try to integrate and test them for hopefully better results.

\subsection{Dividing the work}
The main problem of most of the clustering algorithms (K-means, Expectation Maximization, Hierarchical algorithms, etc.)  is their complexity. With growing number of clusters in the data, most of these algorithms run very long and, specifically for our data, tend to join more clusters together. With an average number of one hundred points per one cluster, we have cca four hundred million clusters in our dataset. None of the more complex clustering algorithms can process such data with a reasonable result - at least not as whole.

So the key here is to divide the data into small chunks not only for helping with the complexity, but also for improving the actual results of the clustering algorithms. When each instance of the clustering algorithm runs for a data set with 5 clusters, each with 100-1000 points, it still provides very good results. It would be even better, if we could be sure, that each one of these chunks actually contains all of the points of it's clusters, that we cannot find any sliced cluster on the edges of these.

There is a quite simple solution because of one fact we know of our data. The actual errors in the observation coordinates come from the inaccuracy of astrometry computed for these objects defined in chapter~\ref{secOurSolution}, i.e., the size of clusters is smaller than this inaccuracy. This technique is the same for all of our data, so we can define a threshold which is greater than the known maxium size of a cluster and use it to define overlaps at the edges of our chunks, which will be wider than this threshold. That ensures with one hundred percent assurance, that we cannot slice a cluster. We will show this in the Realization chapter in more detail in chapter~\ref{chapRealisation}.

\subsection {GPU computing}
There is also the possibility to accelerate the computational process by moving the most expensive processing to GPUs instead of processor. However, to be this enhancement really efficient, this requires very intensive memory micro-management. Also, the advantage of GPUs originates from the ability to process high amount of small tasks which require very little communication with the memory. The clustering algorithm has to be deeply analyzed and bottlenecks quite different from standard processor computing have to be identified and addressed. 

The GPU computing can be actually implemented in an existing C or C++ application, transforming only small parts of the program, where we know from the testing it is efficient. This is possible for example with an nVidia CUDA Toolkit~\cite{cuda}. This toolkit of libraries can be called as simple C++ functions.

There are several solutions implementing K-means using a GPU acceleration. We are planning to take a little different approach, where we want to decrease the number of points for one K-means run and parallelize these tasks. Parallelization inside these tasks would not be that efficient, as it is better if we effectively decrease the complexity by assigning less points per one K-means instance, than to parallelize the one K-means instance for a bigger data set.

After considering all of the above mentioned facts we decided against implementing GPU optimization in our application.

\chapter{Design}
All of the aspects of possible solutions and their results, mentioned in the previous chapter, were carefully considered and a final design decision was made. The accepted and fully implemented solution will be described in this chapter.

We will implement the catalog generation as a stand alone C++ application. This application will take data from PostgreSQL, produce results, and these will be transfered back to the database. With this approach we don't have to change the underlying architecture of the data manipulation at all.

\section{MPI vs OpenMP}
The dimensionality of our data is not very high (2 coordinates, photon flux or magnitude possible as 3rd dimension), so we can load the whole data into memory and process it there. There is a possibility of streaming the intermediate parallel task distributions on disk, but the problem is that we cannot be sure that a task is complete before we iterate over all points in the data set.

The processing cost for the individual tasks will be small in the terms of memory. It depends on the task size, but compared to memory amount needed for constructing these tasks, it is very small. The benefit of distributing the work over multiple nodes is not high, as the tasks are supposed to be small. Another reason why we'd like to try the threads approach is that it is easier and faster to implement. And the easies way to implement a thread-based producer consumer application is OpenMP library~\cite{openMP}, so that is our final decision for parallelization.

The parallel tasks will be processed by an arbitrary number of threads and the results collected by a master thread and written back to disk.

\section{Design details}
\begin{figure}[h!]
\label{figPipeline}
  \centering
    \includegraphics[width=1\textwidth]{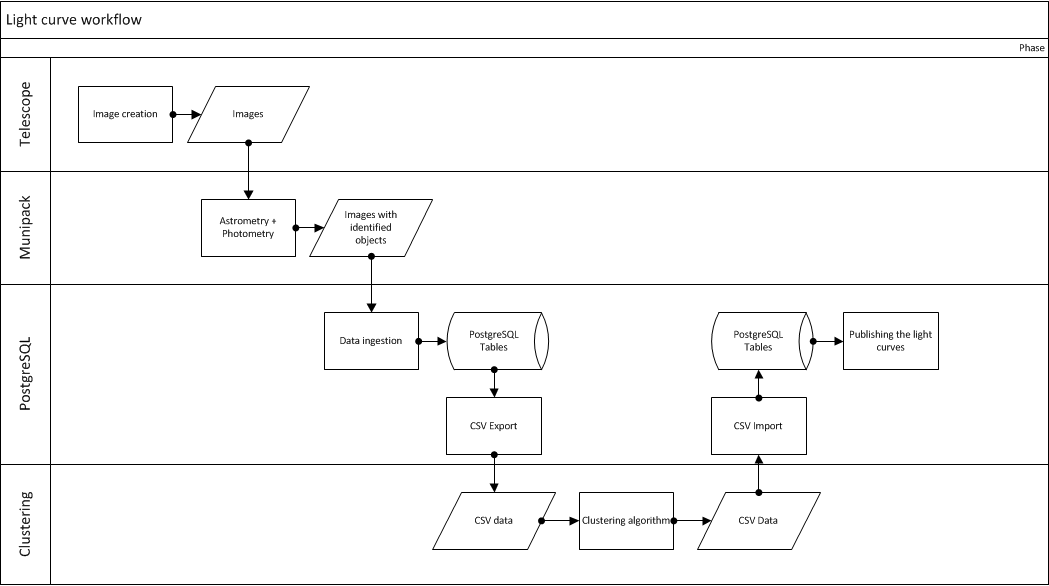}
  \caption{Light curve generation workflow}
  \label{figurePipeline}
\end{figure}

We can see the complete pipeline used for the light curve generation on Fig.~\ref{figPipeline}. The main part of our thesis is  about the Clustering swim-lane.

\subsection{Clustering algorithms}
We also had to analyze multiple clustering algorithms and choose which ones will be used within the C++ application. The possibilities here were a simple K-means, Expectation maximization based techniques such as Gaussian mixtures, agglomerative or divisive hierarchical clustering or last but not least some clustering algorithm of our own. 

We also considered integrating R~\cite{rLanguage} with the Seamless R and C++~Integration~\cite{rcpp}. There is a very large variety of clustering algorithms implemented in R. However, most of these are usable only for testing  on small amounts of data, their complexity grows very quick. We also don't have any possibility to influence the behavior of these R modules, they have to be taken as-is. 

Final reason why we favored C++ library over the integrating R modules was the usability for big amounts of data. C++ is way faster than R in equal computations and has the advantage, that we can debug and potentially change or even repair the library we are using.

In the end, we decided to implement the \emph{Iterative query} algorithm defined in chapter~\ref{secIncrementalQuery}, which will be used as a benchmark to other clustering algorithms. Then we decided to use a library for randomized K-means algorithm~\cite{kmlocal} and the Armadillo C++ library~\cite{armadillo}for EM algorithm based on Mahalanobis distance.

\subsection{HEALPix~\cite{HEALPix}}
\label{secHEALPix}
We also had to consider which sphere indexing library will we use to work with the spatial data. This cannot be done without a library because we need a complex way of building the chunks used for parallel processing. Another reason is a pre-implemented optimized way of computing angular distance between two points on a sphere.

The possibilities here were HTM indexing algorithm~\cite{htm}, HEALPix or Quad Tree cube algorithm~\cite{q3c}. We will not do a detailed benchmark here, as all of these algorithms are quite effective. The decision was simple in the end.

We chose HEALPix because of these major reasons:
\begin {enumerate}
\item HTM~\cite{htm} has higher complexity for higher resolutions of the sphere tessellation, whereas HEALPix~\cite{HEALPix} has constant complexity in this direction. We will work generally only with the high resolutions with the clustering algorithm. The pixels have different sizes at different areas of the sphere.
\item Quad Tree Cube~\cite{q3c} will also produce different sizes of the individual pixels, thus it would be more complicated to distribute the data into chunks uniformly.
\item HEALPix~\cite{HEALPix} can distribute the data uniformly, because it's pixels are of equal sizes on the same resolution. It is also very popular lately and has been implemented into Aladin VO~\cite{aladin} client, which makes it much easier to test and compare our results.
\end {enumerate}

\section{Functional requirements}
We can summarize the results of our analysis to the following requirements on our application.

\begin{enumerate}
\item Data input will be processed from CSV format. This input file will have 4 columns identifying an observation. The \emph{obsnameID} and \emph{starNo} are used as a composite primary key for the observation table. The exported columns will be the following:
	\begin{enumerate}
	\item \emph{Right ascension}
	\item \emph{Declination}
	\item \emph{imageID}
	\item \emph{starNo}
	\end {enumerate}
\item Data ouptut will be again in CSV format and will take place in two files. First one will represent the catalog with following columns:
	\begin{enumerate}
	\item \emph{Catalog ID}
	\item \emph{Right ascension}
	\item \emph{Declination}
	\end {enumerate}
And the second one will represent the observation mappings to these catalog files, having the following columns. Again, \emph{obsnameID} and \emph{starNo} are used to identify observation in database. The columns defining our catalog are:
	\begin{enumerate}
	\item \emph{Catalog ID}
	\item \emph{imageID}
	\item \emph{starNo}
	\end {enumerate}
\item The program will be able to create a groups of the observations, which can be processed in parallel completely separately without degrading the results. This will be accomplished via the HEALPix library~\cite{HEALPix}.
\item It will support several clustering strategies for comparison. Specifically these will be:
	\begin{enumerate}
	\item Our own incremental strategy described in~\ref{secIncrementalStrategy}
	\item Several variants of K-means from KMLocal library~\cite{kmlocal}.
	\item Expectation Maximisation strategy from Armadillo C++ library~\cite{armadillo}.
	\end {enumerate}
\item The solution will be integrated into our current data center. This integration will be implemented by scripts for exporting and importing CSV data for PostgreSQL database.
\end{enumerate}

\section{Application model}
\label{sec:design}
\begin{figure}[h!]
  \centering
    \includegraphics[width=1\textwidth]{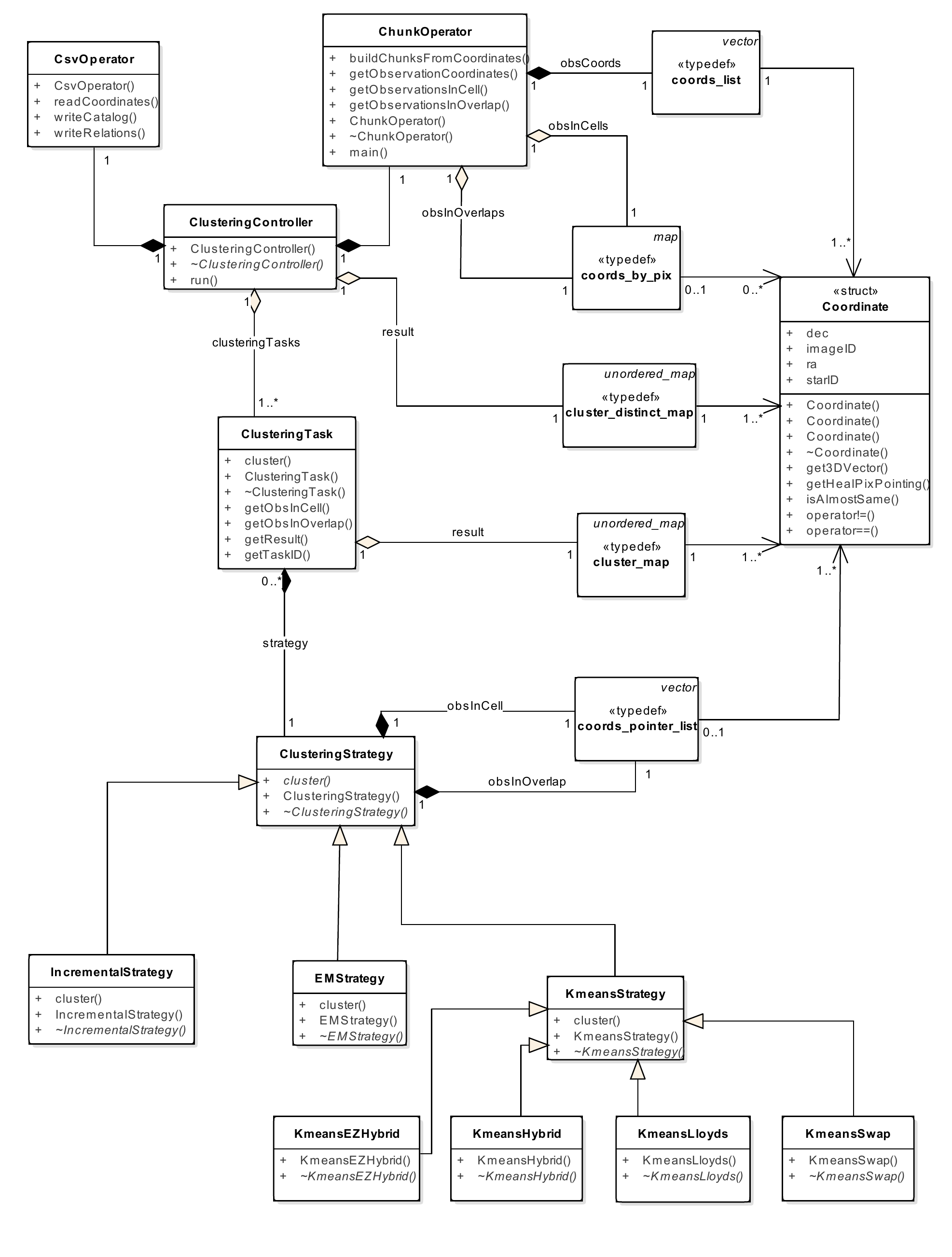}
  \caption{C++ application UML class diagram}
  \label{figureUML}
\end{figure}

The UML diagram for our application can be seen on Fig.~\ref{figureUML}. We will explain the design on a typical run of our application.

The main part of the application logic is contained in class \emph{ClusteringController}. The entry point function just parses the command line arguments and passes them to the the controller. Then it calls the run function. The \emph{ClusteringController} class holds instances of two other controllers - the \emph{CsvOperator} and \emph{ChunkOperator}. 

\emph{CsvOperator} is responsible for working with input and output CSV files. The \emph{CsvOperator} will parse the input file into a vector of \emph{Coordinates} and pass it to the \emph{ClusteringController}. This vector is logically owned by \emph{ChunkOperator} which works with this data most as with \emph{obsCoords} vector. 

Then, the \emph{buildChunksFromCoordinates} method is called on \emph{ChunkOperator}, and the result is stored in two maps - the \emph{obsInCells} and \emph{obsInOverlaps}. In these maps, \emph{Coordinates} are assigned to their HEALPix~\cite{HEALPix} pixel IDs of selected resolution, along with their overlaps if there are any. This function is described more in chapter~\ref{secImplBuildChunks}.

For each of these HEALPix pixels, a \emph{ClusteringTask is created}. This clustering task has a strategy by which it is to be solved. Each strategy is holding a list of pointers to the \emph{Coordinates} it has to process. The \emph{ClusteringController} ensures that tasks are processed in parallel by a number of threads, which has been allocated to this run of our program. Each clustering task will have it's own results stored in \emph{cluster\_map} type, which is a map of \emph{clusterIDs} pointing to a list of it's members. The \emph{clusterID} is too an instance of \emph{Coordinate} struct representing the cluster medoid.

These individual task's results have to be collected into an overall result of the clustering algorithm. This is done sequentially after the parallel phase of actual clustering has been finished. We iterate over the tasks, throw out the duplicate results of neighboring tasks, or merge them into common results. 

The overall result is kept in the \emph{ClusteringController's} result, where we ensure, that a cluster will not have duplicate members, i.e., the list of cluster members is a set of \emph{Coordinates}.

\subsection{Configuration}
The application will take the following configuration. The values values here are optimal according to the testing results.  The rest of configuration, such as K-means configuration, will be listed later.
\begin{lstlisting}
[parallelOptions]
bigPixelNsideExp=15		;Parallel task resolution
overlapPixelNsideExp=18		;Overlap pixel resolution

[resultOptions]
IDNsideExp = 29			;Catalog ID generation resolution
clusterDuplicatesArcSec = 0.5	;Distance at which two clusters will be identified as duplicates <arcsec>

[incrementalStrategy]
catalogIndexNsideExp = 17	;Resolution of the index inside one task
clusterRadiusArcSec = 1 		;Distance at which I will add a point to the cluster <arcsec>
\end{lstlisting}
\newpage

\section{Time Complexity}
The most crucial application phases have the following complexity:
\begin{enumerate}
	\item{Building chunks} - This phase has a \(O(n)\) complexity, as it touches every observation once.
	\item{Clustering phase} - Dependent on a clustering algorithm, this is either \(O(n)\) for a simple strategy, or \(O(n^2)\) for other strategies implemented (K-means, EM)
	\item{Collecting results} - \(O(n^2)\) for merging the results. Depending on the task size, we have to check the neighboring tasks for duplicate clusters and in worst case, merge them all.
\end{enumerate}

\section {Scalability}
The scalability of our solution results from the chosen architecture. As we are using an in-memory solution based on threads rather than individual processes, we are favoring a shared-memory architecture. The  memory limitations for the whole application run are around 8~GB per 100 million points. This relation is linear, so for 1 billion points we would need an 80~GB RAM machine with one processor to do the division into parallel tasks. These individual task's complexities depend on the strategy used for solving these tasks, but the idea is to keep them small for better efficiency and result quality of the clustering algorithms beneath.

\chapter{Realisation}
\label{chapRealisation}
As we already deduced above, the most appropriate solution to our problem is a stand alone C++ program, which will be integrated into our current data center. We will comment the implementation in this chapter, pointing out the interesting parts, which solved the problems the previously mentioned solutions could not. We will also document here the way of integration with our current solution of the original data ingesting and publishing the light curves.

\section{Implementation}
In this chapter we will present the interesting parts of our solution, with examples of C++ code.

\subsection{HEALPix usage}
\label{secHEALPixUsage}
We are using the HEALPix library~\cite{HEALPix} to create the parallel tasks, which can be processed individually, without need of further communication between themselves. The key aspect for defining such tasks is the \emph{Nside} parameter of the HEALPix grid. The \emph{NSide} is specifying the size of HEALPix pixel. The number is always \(2^N\), where N can range from 1 to 29. For our case, the most convenient are around 18, which specifies a pixel approximately the size of one cluster. We use these pixels for defining the overlapping the region of our parallel tasks, which are equivalent to the area of a HEALPix pixel of a \(Nside = 18\). The \emph{Nside} used for the parallel tasks is meaningful between 10 and 15 (for 16 and more the overlap is actually bigger than the task area itself).

For simplicity, the \emph{"task size"} term will be used for the exponent of an \emph{Nside}. For \(Nside 2^15\), the \emph{task size} will be 15, for \(2^29\) the \emph{task size} will be 29. Whenever we will talk about the \emph{task size}, we will refer to this exponent, not the actual \emph{Nside} value.

\subsection{Creating spatial chunks}
\label{secImplBuildChunks}
This section is describing functionality in \emph{ChunkOperator's buildChunksFromCoordinates} function. We are mentioning here the most interesting parts of the code. For the whole function, you can look at appendix~\ref{codeChunks}.

First we transform our coordinates to the system HEALPix~\cite{HEALPix} is using and then comes the interesting part. We create two HEALPix bases, first one with the resolution of the task spatial partitioning (i.e. How big will be the area for one individual task) and the second one with the resolution of the overlaps.
\begin{lstlisting}
HEALPix_Base2 base1(base1Nside, NEST, SET_NSIDE);
HEALPix_Base2 base2(base2Nside, NEST, SET_NSIDE);
\end{lstlisting}

Then we iterate over observations and decide in which cell it lies.

\begin{lstlisting}
int64 idx_lores = base1.ang2pix(observations[i]);
\end{lstlisting}

Then we check, whether this point does not belong to an overlap (based on the base2 finer grid) of the neighboring cell (based on the base1 rougher grid). The big cell (i.e. task) can have up to 8 neighbors in the HEALPix geometry.
\begin{lstlisting}
// now check whether the surroundings of the observation touch neighboring jobs
        int64 idx_hires = base2.ang2pix(observations[i]);

        fix_arr<int64, 8> neighbors;
        base2.neighbors(idx_hires, neighbors);
\end{lstlisting}
We iterate over these 8 neighbors and if it happens that the \emph{nbidx\_lores} (i.e. big pixel ID next to our small pixel used for overlaps) is not actually the current \emph{idx\_lores} then we add our current point to this neighboring big pixel's overlapping region.
\begin{lstlisting}
int64 nbidx_lores = base1.ang2pix(base2.pix2ang(neighbors[j]));

if (nbidx_lores != idx_lores) { // touches a neighbour cell  
  if ((*obsInOverlap)[nbidx_lores].empty() ||
          (*obsInOverlap)[nbidx_lores].back() != &(*obsCoords)[i]) {
         (*obsInOverlap)[nbidx_lores].push_back(&(*obsCoords)[i]);
   }
}
\end{lstlisting}
By this algorithm we can distribute any kind of spherical coordinates into chunks of equal size (equal area on the sphere) with overlaps of variable size in a linear complexity (touching each observation only once). This was the biggest obstacle in parallel clustering of our data.

\subsection{Actual clustering}
Each task remembers the strategy\footnote{Strategy is a software design pattern used for solving same task by different ways of doing it} by which it should be solved.

The parallel processing of each task is resolved by OpenMP library~\cite{openMP}. The clustering loop looks like this. The \emph{omp\_set\_dynamic(0)} is used to enforce that the number of threads we provide will be actually used for the computing. 
\begin{lstlisting}
omp_set_dynamic(0); // Explicitly disable dynamic teams
omp_set_num_threads(this->noThreads); 
#pragma omp parallel for if (this->noThreads > 1)
    for (size_t i = 0; i < clusteringTasks.size(); i++) {
      ClusteringTask *currTask = clusteringTasks[i];
      currTask->cluster();
}
\end{lstlisting}

\subsubsection{Incremental strategy}
\label{secIncrementalStrategy}
We wrote this strategy as a benchmark for the other strategies such as K-means. Those were taken from 3rd party libraries mostly, so we will not document them here.

This strategy is basically re-written \emph{Incremental query} from chapter~\ref{secIncrementalQuery}. We iterate over observations in our cell, then we iterate over the ones in our overlap in the same way and finally remove clusters we decide to be incomplete (i.e. they were sliced on the edges of the overlap region). We remove the incomplete clusters when using other strategies too.
\begin{lstlisting}
void IncrementalStrategy::cluster(cluster_map &taskResult) {
    processObservations(obsInCell, taskResult);
    processObservations(obsInOverlap, taskResult);
    removeIncompleteClusters(taskResult);
}
\end{lstlisting}
The code for processing each individual observation is simple too. For each observation, we check whether we don't have a catalog ID close to it already. If there is one, we just update it and add ourselves to that catalog ID. This is done in \emph{findAndUpdateNeighbor} function. 

If we could not update the catalog, that means we have to create a new identifier with the same coordinates as the current observation. We set the \emph{imageID} and \emph{starID} to -1 as this observation is not originally from database. The \emph{catalogIndex} is used for indexing already processed observations of this task in memory for faster neighbor lookup.
\begin{lstlisting}
void IncrementalStrategy::processObservations(coords_pointer_list *obsList,
        cluster_map &taskResult) {
    for (coords_list_it it = obsList->begin(); it != obsList->end(); it++) {
        Coordinate * currObs = *it;
        int64 indexID = indexBase.ang2pix(currObs->getHEALPixPointing());

        bool catalogUpdated = false;
        catalogUpdated = findAndUpdateNeighbor(indexID, currObs, taskResult);

        if (!catalogUpdated) {
            Coordinate * clusterID = new Coordinate(currObs->ra, currObs->dec, -1, -1);
            taskResult[clusterID].push_back(currObs);
            catalogIndex[indexID].push_back(clusterID);
        }
    }
}
\end{lstlisting}

The \emph{findAndUpdateNeighbor} method gets close points from the \emph{catalogIndex} (it does not search all of the points, only the ones in neighboring pixels). Then it iterates over these coordinates and for each one computes distance to it. If the distance is smaller than a configured value, it is added to that cluster. If not, we return false and a new cluster is based on it.
\begin{lstlisting}
bool IncrementalStrategy::findAndUpdateNeighbor(int64 indexID,
        Coordinate * currObs, cluster_map &taskResult) {
    coords_pointer_list clustersCloseToMe;
    getClosePoints(clustersCloseToMe, indexID);

    for (coords_list_it it = clustersCloseToMe.begin(); it != clustersCloseToMe.end();
            it++) {
        Coordinate * currNeighborClusterID = *it;
        double distance = HEALPixHelper::computeDistance(currObs, currNeighborClusterID);
        if (distance < Config::clusterRadius) {
            updateNeighboringCluster(currNeighborClusterID, currObs, taskResult);
            return true;
        }
    }
    return false;
}
\end{lstlisting}

\subsubsection{K-means strategy}
We used the K-means algorithm from library KMlocal library~\cite{openMP}. The metric used for optimization here is the Euclidean distance of our coordinates to the K-means centroids called distortion. We are using average distortion for better compatibility with the elbow method. Along with minimizing this function, we also try to minimize the number of clusters. The algorithm is enhanced with simmulated annealing. The randomization it brings gives us far better results with the right parameters, as it greatly reduces the risk of getting stuck in a local minimum.

A K-means algorithm has to have a parameter K - the number of clusters. We don't have this number, however, and need to estimate it. The elbow method is a standard method of choosing the right K. It is based on iterating the algorithm for different K parameters and choosing the one with the best ratio of minimizing K as well as the average distortion.

As our data is forming clusters of approximately the same size, we can get away with a simple identification of the elbow in our graph of average distortions, using just the ratio of average distortion of previous K to our current K.

\begin{figure}[h!]
  \centering
    \includegraphics[width=1\textwidth]{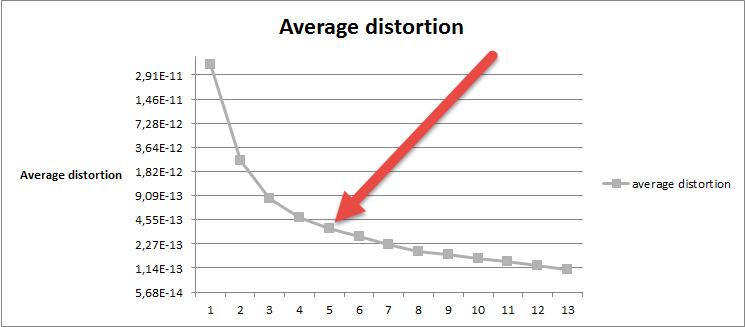}
  \caption{Elbow method}
  \label{figureElbow}
\end{figure}

\subsubsection{Parameters}
\begin{lstlisting}
[K-meansLocalStrategy]
maxClusters = 10000		;maximum number of tested clusters per 1 task	
maxTotStageVec0 = 100	
maxTotStageVec1 = 10
maxTotStageVec2 = 2	
maxTotStageVec3 = 1		;number of stages = a + (b*k + c*n)^d
minConsecRDL= 0.10		;min consec RDL
minAccumRDL	= 0.10 	;min accum RDL
maxRunStage = 3			;max run Stages
initProbAccept = 0.5		;init probability of acceptance
tempRunLengt = 10 		;temp. run length
tempReducFact = 0.95		;temp. reduction factor
elbowFact = 2 			;elbow Method aceptance factor
\end{lstlisting}

The number of maxClusters is not that important, as the \emph{task sizes} for the K-means should be much smaller than this number, where the elbow method will just choose the right K and end the computation. If we want to use bigger \emph{task sizes} with thousands of clusters, this algorithm will run much slower, but it will run nonetheless\footnote{The problem will be with the \emph{elbowFactor}, as it is tuned for the \emph{task size} 15. For lower \emph{task sizes} (i.e. bigger amount of observations), it can converge more slowly and for big K the algorithm will run very long}. The \(O(n^2)\) complexity for each individual task depends on the actual best K for that cluster according to our elbow method, not on the maxClusters parameter.

Most important is the number of stages (iterations) for the whole clustering algorithms. This represents the total number of randomized starts of the algorithm to get the best result possible. Choose this too small and the quality will drop off, choose this too high and the clustering will be very slow. The complexity for each task is O(n*m), where n is the number of running stages and m the actual complexity of the individual K-means run.

The elbowFact is the coefficient of \begin{math}d_{K-1}/d_{K}\end{math}, where K is the actual tested number of clusters and d the average distortion, which needs to be satisfied to accept the solution of K as best known so far. Otherwise, the algorithm terminates.

The other parameters are specific for the different kinds of K-means algorithm used and will be discussed in the result chapter~\ref{chapResult}.

\subsection{Collecting results}
When all of the tasks are complete, we need to collect the results from each individual task and merge them into a common result. Thanks to the overlaps, the clusters produced by individual tasks will never be incomplete (those we threw away), but instead, there will be duplicate identifications of the same cluster in spatially neighboring tasks.

Based on the clustering technique used, these clusters will be exactly the same, or will be close to each other. As we can compare cluster centers only based on their floating point coordinates, we cannot rely on exact equality. Instead, we need to define a threshold distance at which we will declare two clusters to be duplicates and need to be merged.

Because we already know the accuracy of our astrometry, we can even improve the results of clustering algorithms inside the individual tasks. For example K-means tends to divide clusters in particularly dense areas into multiple smaller ones. These have their centers very close to each other (typically less then 500 mas). 

The merging technique will serve two purposes then:
\begin{enumerate}
\item Remove duplicit results when collecting results on the edges of individual tasks.
\item Merge duplicit results of clustering inside the tasks.
\end{enumerate}

Thus, for each clusterID, we try to find other cluster centers closer than a given threshold (in our case cca 500mas) and from these choose the closest one and merge with it. The coordinates of a resulting cluster are re-computed by weighted coordinates of both of the merged clusters (based on number of members in each cluster). For merging the actual points we need to have a set container for the resulting cluster members, which causes additional, but inevitable memory overhead. More about the memory topic in chapter~\ref{subSecMemory}.

This technique of merging can be described as a simple 1 step streaming K-means algorithm. It will still not ensure that the corner cases such as the results of underlying clustering being in one line will be solved. But as the points entering this phase should be already accurate results of the underlying clustering algorithm, these cases will be extremely rare.

\subsection{Memory optimization}
\label{subSecMemory}
The critical memory consumption point in our program is the distribution of our data into individual tasks. As only the majority (but not all) of our data is spatially localized (sorted), we have to go through the whole dataset to be sure that each task is complete (it contains all points needed to process the task without further communication). 

One observation can be represented by two doubles (coordinates) and two Integers (database ID), which is 24B in memory. For our whole dataset, this stands for \(4 * 10^8 * 24B = 8GB\) . This is still pretty usable on shared memory based architecture. We can still reduce the memory by using some kind of streaming for this process, but we postponed this optimization for later versions. 

Then we start building the pointers to these coordinates which represent processing tasks at first, then clustering groups.

At the end, we need to collect the results. This process is the most memory consuming, as we have to ensure the clusters to be without duplicates. This is ensured by using a set container based on a tree structure - effectively using much more memory per one pointer, than a simple vector.

\begin{figure}[h!]
  \centering
    \includegraphics[width=1\textwidth]{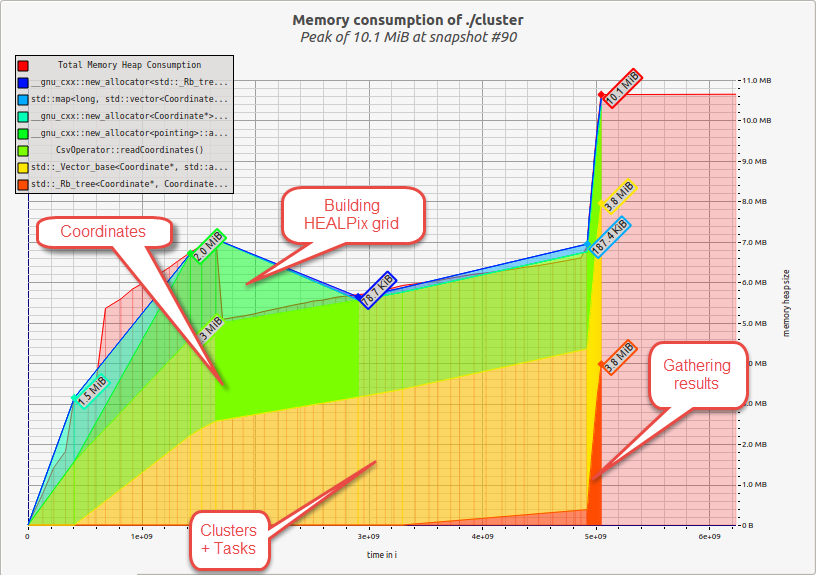}
  \caption{Basic memory consumption}
  \label{figMemoryBase}
\end{figure}

\begin{figure}[h!]
  \centering
    \includegraphics[width=0.9\textwidth]{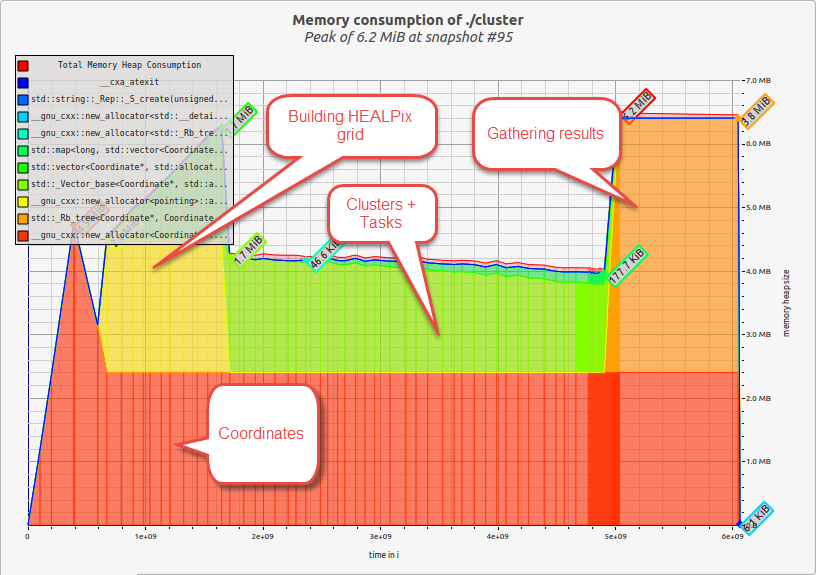}
  \caption{Memory savings applied}
  \label{figMemoryBest}
\end{figure}

With simple memory usage, we can see the memory consumption on a small data set of one hundred thousand observations without optimization on Fig.~\ref{figMemoryBase} and with optimization on Fig.~\ref{figMemoryBest}. These graphs were taken with Valgrind tool Massif~\cite{massif} and displayed with Massif Visualizer~\cite{massifVisualizer}.

The most efficient savings come from shrinking the data vector capacity when all coordinates are loaded and from discarding coordinate pointers to tasks that are already processed. Another enhancement that could be made is using integer indices to the data vector, instead of using direct pointers to it's contents (8B for a pointer vs 4B for an array index - we don't have more than integer size observations). This micro-optimization will be implemented if needed, but is not part of our current solution.

At the end of the program run, we can see a fast grow in memory. This is consumed by using a set to store the clustering results for each cluster. This is actually needed to ensure we will not have duplicate entries in one cluster after merging it with another one. At the end of the program, we have only this result map and the original data structures in memory, no more memory can be saved.

\section{Integration}
The integration with PostgreSQL data is very simple, based on CSV export of information required for the actual clustering, and CSV import for the results of clustering. The tables used for export and import can be seen in chapter~\ref{secDBBack}.

\begin{figure}[h!]
  \centering
    \includegraphics[width=0.9\textwidth]{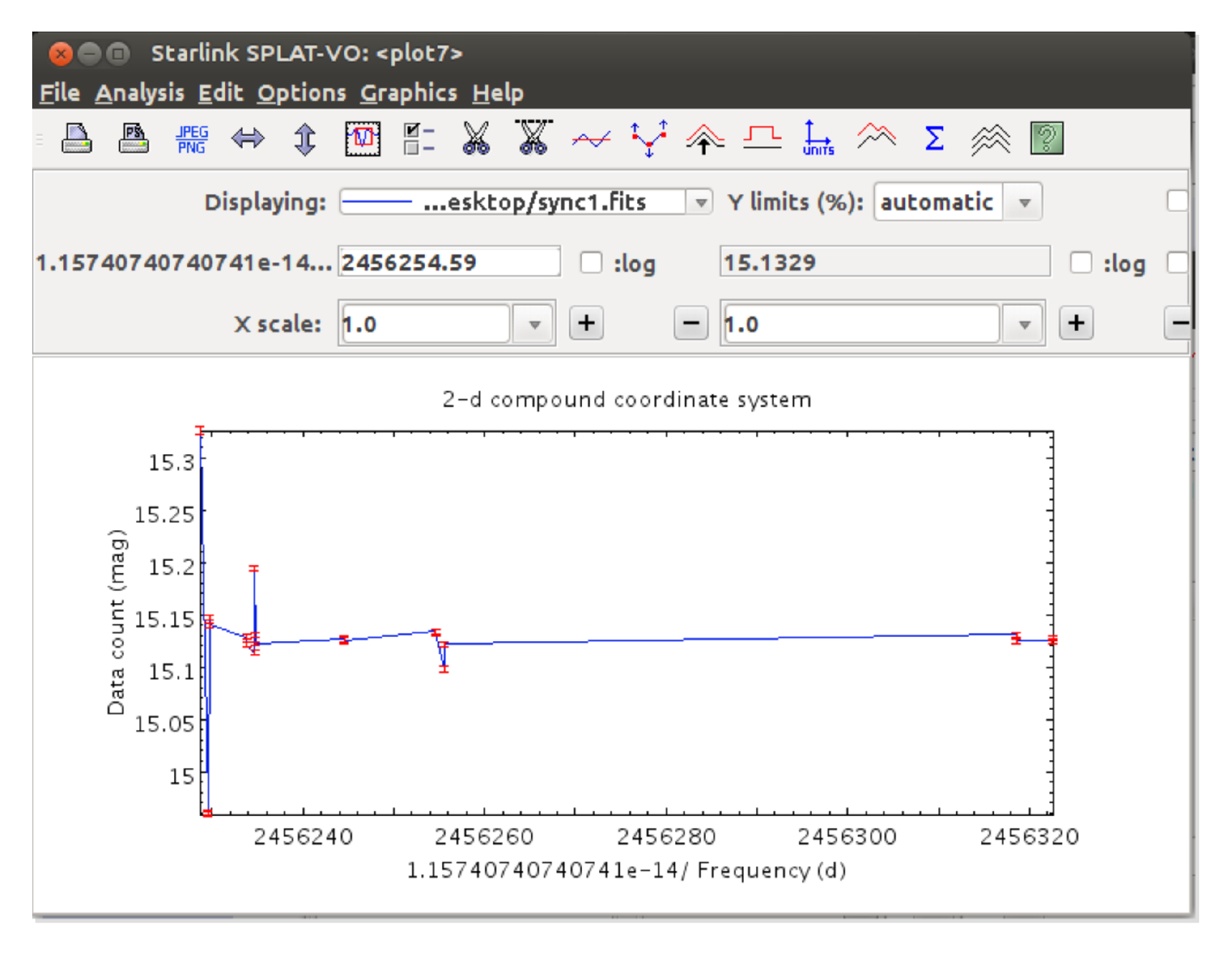}
  \caption{Resulting light curve displayed in SPLAT-VO}
  \label{figSplat}
\end{figure}

The result of our work can be then published by the mechanism already implemented in my Bachelor's thesis~\cite{bakule}. The resulting light curve can be seen on Fig.~\ref{figSplat}, displayed in SPLAT-VO~\cite{splat}. The brightness of our star here is around 15 magnitude and the period where the star was observed is from 2456220 Julian Date (19.10.2012) to 2456320 Julian Date (27.1.2013).

\chapter{Results}
\label{chapResult}
We will be measuring our results in both performance and quality. For performance testing we are using a machine with 12 cores, 32~GB of RAM. For quality testing, we will be using various comparison of our results with on-line catalogs as well as analysing the results amongs themselves. We decided to do the comparisons for data on a logarithmic scale, using testing data set composed of 1~million, 10~million and 100~million observations.

The metric of quality will be the fitness of chosen cluster centers (i.e. that the cluster centers were assigned correctly). This can be expressed in various ways and we decided to use the one native to K-means algorithm - the distance of points to their cluster centers (this metric is linearly dependent on distortion used for K-means algorithm).

\section{Time complexity}
Here we will discuss the time complexity of the our algorithm.  The key parameters on which the complexity depends are:

\begin{enumerate}
\item Strategy used
\item Data size
\item Parallel task size
\item Number of threads used
\item Other parameters specific for the strategy used
\end{enumerate}

The strategies measured will be the incremental one described in section~\ref{secIncrementalStrategy} and the hybrid version of K-means~\cite{kmlocal} as it provides the most quality results, as can be seen on Fig.~\ref{figK-meansTest}.

Data sizes used will be as already mentioned 1 million, 10 million and 100 million observations.

The parallel \emph{task size} is explained here~\ref{secHEALPix}.

First we start with the simple incremental clustering strategy.

\subsection{Incremental strategy}
The incremental clustering strategy is described in chapter~\ref{secIncrementalStrategy}. We will analyze the time usage of this simple algorithm for different data sizes.

\subsubsection{Task size based on data size}
\begin{figure}[h!]
  \centering
    \includegraphics[width=0.8\textwidth]{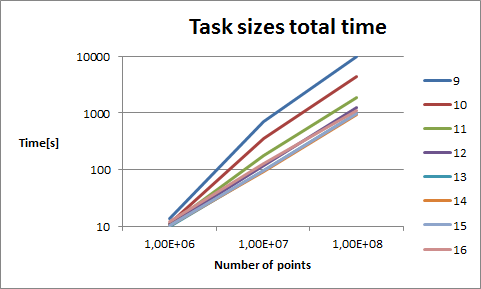}
  \caption{Total time based on data size}
  \label{figIncrementalTaskSizeDataSize}
\end{figure}

On the Fig.~\ref{figIncrementalTaskSizeDataSize} we can see the total time of our clustering algorithm for 1 million, 10 million and 100 million observations. Here we can see that for the lower \emph{task sizes} (i.e. greater number of observations per one task), are taking more and more time. This behavior is explained on Fig.~\ref{figProgramParts}.

\subsubsection{Program phases based on task size}
\begin{figure}[h!]
  \centering
    \includegraphics[width=0.8\textwidth]{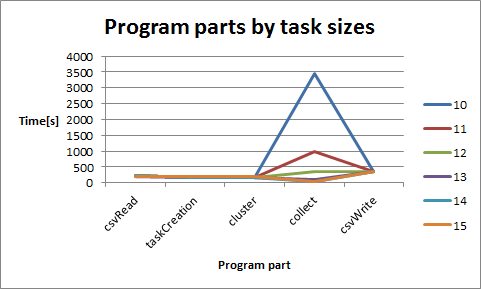}
  \caption{Program parts time}
  \label{figProgramParts}
\end{figure}

On the Fig.~\ref{figProgramParts} we can see running time of parts of the algorithm for 100 million observations based on the \emph{task size}. We can see here, that all of the algorithm parts run the same time, but increasing number of points in one task (decreasing the \emph{task size} parameter) will cause the time of collecting the results. This is because we need to go through more and more results when we are merging the results from individual tasks from the clustering phase.

\subsubsection{Threads based on data size}
\begin{figure}[h!]
  \centering
    \includegraphics[width=0.8\textwidth]{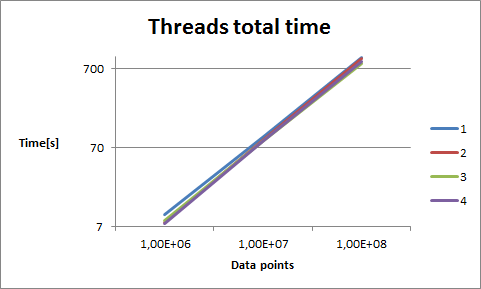}
\caption{Thread time based on data size}
  \label{figThreadsDataSize}
\end{figure}

On the  Fig.~\ref{figThreadsDataSize} we can see here the running time for different numbers of threads for 1 million, 10 million and 100 million observations. The algorithm is linear and increasing number of threads will not accelerate it, because the clustering phase of the algorithm takes in average cca 80\% of the time needed to even read the data from CSV, running on one thread. This relation is compared on Fig.~\ref{figClusterPercent}.

\subsubsection{Threads based on task size}
\begin{figure}[h!]
  \centering
    \includegraphics[width=0.8\textwidth]{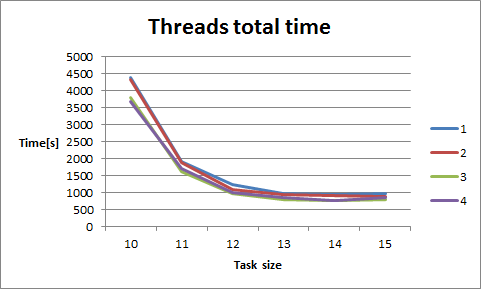}
\caption{Thread time based on task size}
  \label{figThreadsTaskSize}
\end{figure}

On Fig.~\ref{figThreadsTaskSize} we can see running time of different numbers of threads for incremental clustering strategy, 100 million observations based on different parallel task sizes. Again we can see here that the number of threads is not accelerating the computation much, parallelized phase of clustering is too simple and short. For smaller \emph{task size} (bigger amount of objects per one task) the algorithm runs longer.

\subsubsection{Parallel efficiency}
\begin{figure}[h!]
  \centering
    \includegraphics[width=0.8\textwidth]{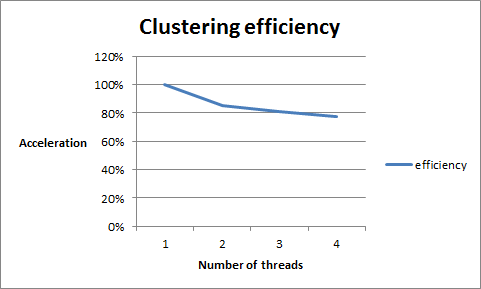}
  \caption{Parallel efficiency for clustering time of linear algorithm.}
  \label{figIncrementalEfficiency}
\end{figure}

On Fig.~\ref{figIncrementalEfficiency} we can see the real time for the clustering phase of our algorithm for 100 million rows. We can see that the parallelization is actually efficient but benefit of parallelization is not worth it as usually the clustering phase takes cca 20 \% of the computational time (it's complexity is linear with the number of points).

\subsection{K-means hybrid strategy}
Because of the higher complexity, we will measure K-means hybrid algorithm with 1 million objects. This number is still considerable when using with this high complexity strategy.

\subsubsection{Clustering relative time}
\begin{figure}[h!]
  \centering
    \includegraphics[width=0.8\textwidth]{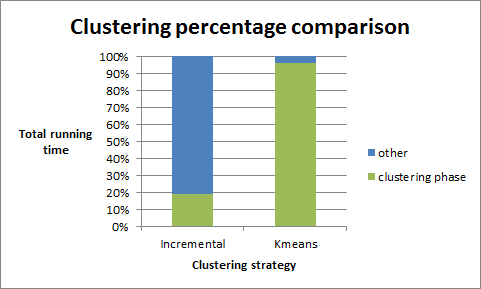}
  \caption{Clustering phase time compared to total running time}
  \label{figClusterPercent}
\end{figure}

On the following graphs we can see the comparison of the clustering phase time and the total time. For incremental algorithm, this is a very low ratio, so the results of parallelisation are poor. For a K-means algorithm, however, the percentage is very high and parallelisation works excellent. The comparison can be seen on Fig.~\ref{figClusterPercent}.

\subsubsection{K-means parallel efficiency}
\begin{figure}[h!]
  \centering
    \includegraphics[width=0.8\textwidth]{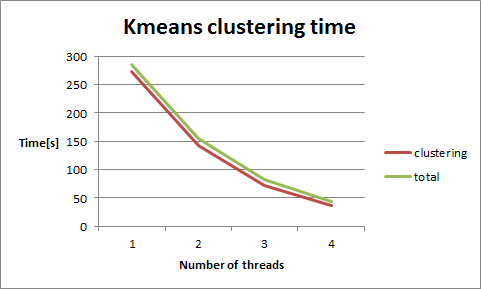}
  \caption{Real time of K-means hybrid algorithm}
  \label{figkmeansThreads}
\end{figure}

\begin{figure}[h!]
  \centering
    \includegraphics[width=0.8\textwidth]{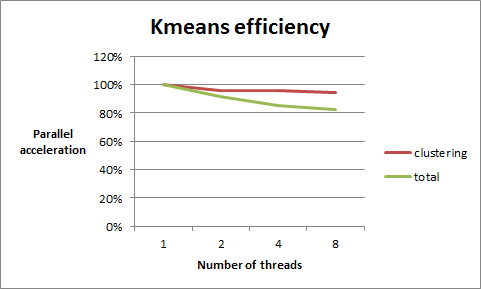}
  \caption{Parallel efficiency fo K-means algorithm}
  \label{figkmeansEfficiency}
\end{figure}

For a smaller data set of 1 million observations, it runs a K-means algorithm for times seen on Fig.~\ref{figkmeansThreads}. The parallel efficiency for this graph can be seen on Fig.~\ref{figkmeansEfficiency}. We can see that the parallelization is really worth it, as the parallel effectiveness is closing to one even for running time under one minute (i.e. the parallel acceleration is close to linear). The results will be even better for bigger data sets.

\section {Quality}
In this chapter we will discuss the results of the clustering algorithms tested. We will discuss the quality of the individual results, as well as their overall error rate and a comparison to on-line catalog.

For most of the result analysis we used the latest version of Aladin~\cite{aladin} with the new functionality of displaying the HEALPix grid. For the statistical analysis such as histograms, we used TOPCAT~\cite{topcat}.

\subsection{HEALPix partitioning}
\begin{figure}[h!]
  \centering
    \includegraphics[width=1\textwidth]{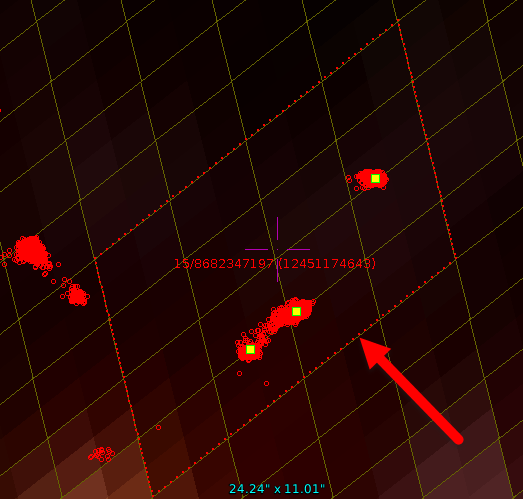}
\caption{Aladin view of the HEALPix grid.}
  \label{figHEALPixExample}
\end{figure}

On Fig.~\ref{figHEALPixExample} the big pixel used for an individual parallel task is marked by the red arrow. The individual observations are marked as red circles, the cluster centers in the current task as yellow squares. The smaller quadrilaterals with yellow borders can be used for the overlapping region around our red marked task, as they are clearly bigger than the cluster size.

\begin{figure}[h!]
  \centering
    \includegraphics[width=1\textwidth]{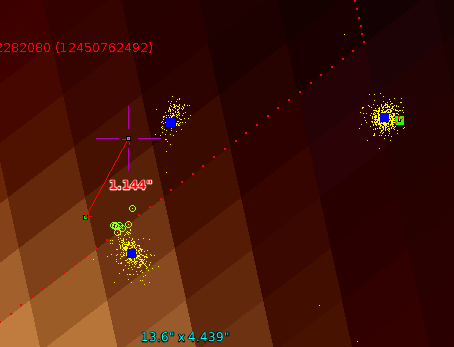}
\caption{Clusters in overlapping region.}
  \label{figHEALPixAladin}
\end{figure}

On Fig.~\ref{figHEALPixAladin} we can see the big pixel used for one task (borders marked by red dots) and the small quadrilaterals seen are used for overlaps. A line of these small pixels along the red line are forming the overlapping region. For the big pixel situated on the bottom of the image we can see the green marked observations are forming a cluster in it's overlapping region and need to be discarded, as they could interact with the cluster in the big pixel above, without having all of the points needed in the same task. 

Another more detailed view can be seen on Fig.~\ref{figHEALPixAladin}

\subsubsection{Expectation maximisation}
\begin{figure}[h!]
  \centering
    \includegraphics[width=0.8\textwidth]{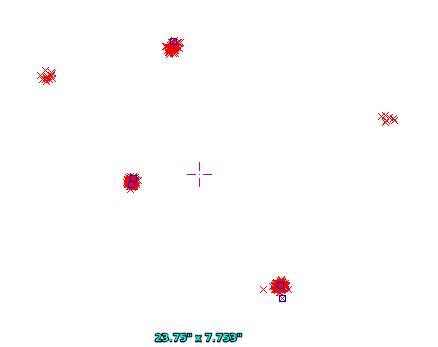}
  \caption{Initial centers chosen by random}
  \label{figEMInit}
\end{figure}

\begin{figure}[h!]
  \centering
    \includegraphics[width=0.8\textwidth]{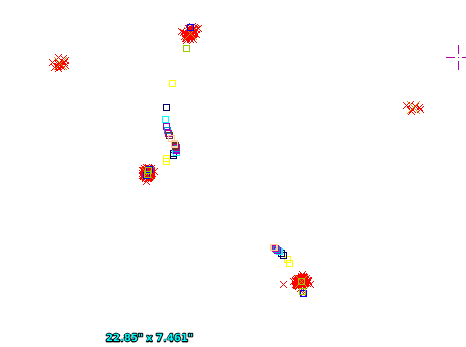}
  \caption{EM convergence}
  \label{figEMProgress}
\end{figure}

Expectation maximization has poor results with our data and we show it on the following images. On Fig.~\ref{figEMInit} we can see the initial centers. On fig.~\ref{figEMProgress} we can see the convergence of these centers during the iterations. No matter what parameters where used, the result always merged all of the data into one big cluster, the differences were only in the speed of convergence to this local minimum.

The reasons can be inappropriacy of the Expectation maximization algorithm for our data, misinterpretation of the arguments, or in the actual implementation we used from a 3rd party library, but we couldn't overcome them and EM algorithm was discarded as inappropriate for clustering our data.

\newpage

\subsection {K-means}
\begin{figure}[h!]
  \centering
    \includegraphics[width=1\textwidth]{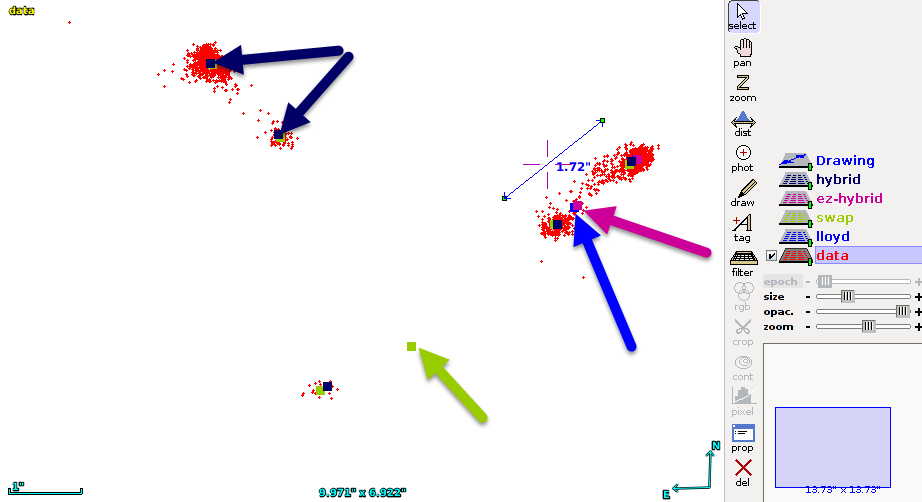}
\caption{K-means variants comparison}
  \label{figK-meansTest}
\end{figure}

Several types of a K-means algorithm are displayed on Fig.~\ref{figK-meansTest}, showing their particular flaws on examples where they misinterpret the clusters. The legend can be seen on the right and the arrows are pointing out the cluster centers this particular version of K-means produced after clustering the data displayed as red dots.

\subsubsection{Elbow factor importance}
\begin{figure}[h!]
  \centering
    \includegraphics[width=0.8\textwidth]{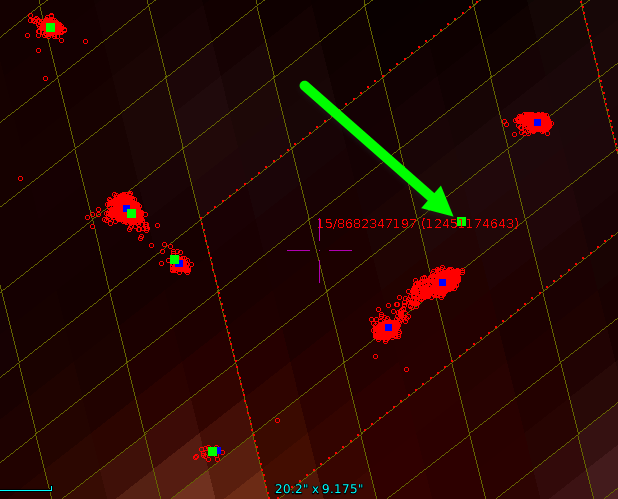}
\caption{Elbow factor too high.}
  \label{figElbow}
\end{figure}

On Fig.~\ref{figElbow} we can see what happens if we specify the elbow factor too high. The willingness of the algorithm to accept higher number of centers is low and it ends before it can separate the data correctly.

\subsubsection{Merging radius too high}
\begin{figure}[h!]
  \centering
    \includegraphics[width=0.8\textwidth]{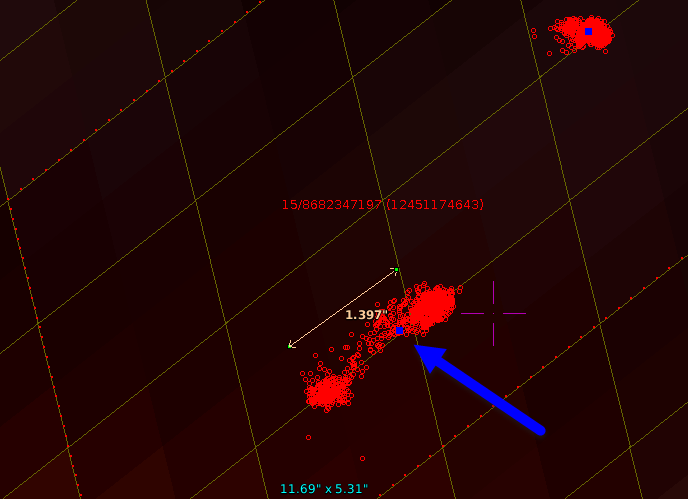}
  \caption{Cluster join radius too high.}
  \label{figJoinRadHigh}
\end{figure}

On Fig.~\ref{figJoinRadHigh} we can see that defining the cluster join radius too high will have similar effect. It joins multiple clusters together even if they were not duplicates.

\subsubsection{Merging radius too low}
\begin{figure}[h!]
  \centering
    \includegraphics[width=0.8\textwidth]{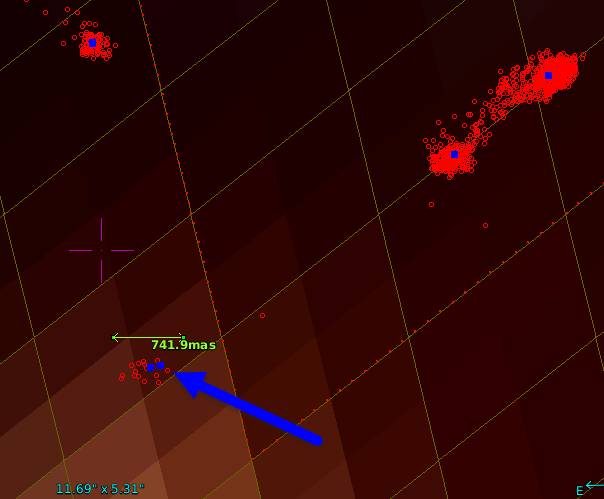}
  \caption{Cluster join radius too low.}
  \label{figJoinRadLow}
\end{figure}

On the other hand, on Fig.~\ref{figJoinRadLow} we can see that specifying the duplicate join radius too low (e.g. 100 mas) will cause the duplicate results on edges of our tasks remain. They will not be marked as duplicates and we result with cluster centers closer to each other than 1 arcsec and with observations split among two duplicate clusters.

\newpage
\newpage

\subsection{Catalog comparison}
Here we will try to cross-match the created catalog with points from which it was created and see the average distance between the cluster center and it's members.

\begin{figure}[h!]
  \centering
    \includegraphics[width=1\textwidth]{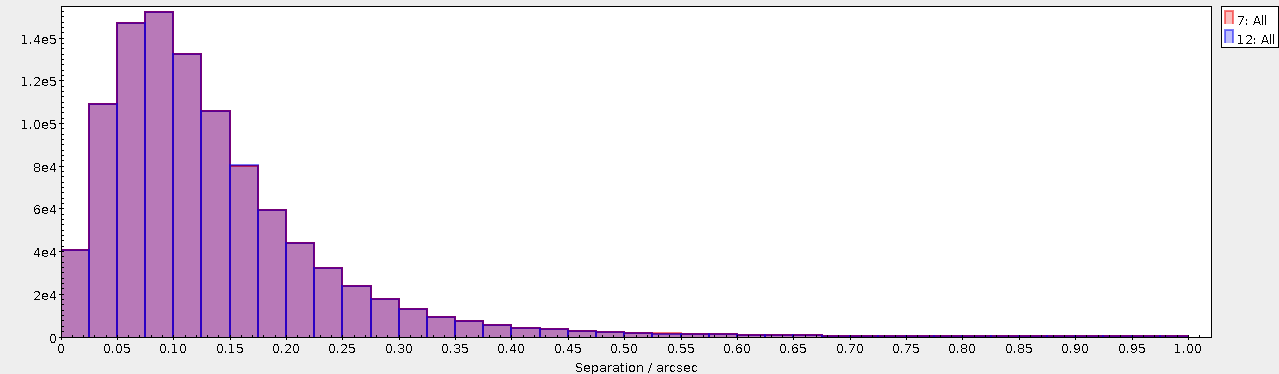}
  \caption{Catalog cross-match for incremental strategy}
  \label{figCatalog}
\end{figure}

The graph on Fig.~\ref{figCatalog} has a pattern of chi-squared distribution. This fact comes from the way we are computing distances. For one dimension, the term can be simplified as \( (x_1 - x_2)^2 \). It is not important, whether \(x_1\) is greater than \(x_2\), this information is lost with powering the subtraction. So the most points on a histogram won't indeed be around zero, but around mean accuracy of the underlying astrometry process, which we stated to be around 0.25 arcsec. 

The catalog identifiers are even more precise than the original astrometry, as they take average coordinates for each cluster, effectively reducing the error. The fact, that we get this histogram with high quality on-line catalog means that we are identifying our cluster centers very precisely, as the mean distance does not shift to higher numbers, but remains below the mean value of the astrometry accuracy.

On Fig.~\ref{figCatalog} we can see the results of our incremental algorithm for one hundred million rows cross-matched to 2MASS on-line catalog. It is a histogram of distances between our catalog identifiers and the ones in the on-line catalog. The red is for using big pixels for parallelization (\emph{task size 10}), the blue is for finer ones (\emph{task size 15}). We can see the results are almost the same - work parallelization is not costing us quality in the case of incremental algorithm.

\begin{figure}[h!]
  \centering
    \includegraphics[width=1\textwidth]{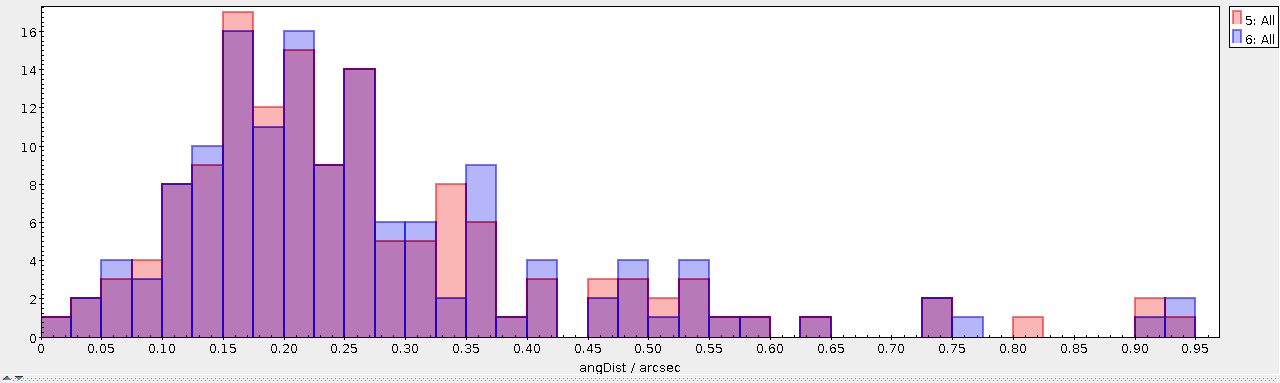}
  \caption{Catalog cross-match for incremental and K-means strategy}
  \label{figKmeansIncrCatalog}
\end{figure}

The results of the k-means algorithm are differing very slightly from our own algorithm and if, they are worse. We can see that on Fig.~\ref{figKmeansIncrCatalog} where catalog generation for 1~million observations is compared. The blue histogram is for distances from K-means centers to SDSS catalog objects and the red one for distances between the incremental strategy centers to SDSS catalog objects.

\section{Result summary}
The result quality of different strategies explained above is highly dependent on the actual data that is fed into the program. For the tested data our own incremental strategy provides more accurate results. For more sparse data, however, we believe that the K-means algorithm will produce better results and here the parallel acceleration of the whole solution will kick in.

Nevertheless, Both of the methods provide high quality results and work exactly as expected in the design phase of our project.

\setsecnumdepth{part}
\chapter{Conclusion}
The goal of this thesis has been met. We finally managed to create a fully operational solution, which can in small time create a catalog of light curves of all our observations. The different approach we chose because of the nature of our data made it very hard, but we succeeded. Our approach can be now reused in any other similar sky survey where the light curve information was never mined from it's data.

In the end, the results of our own clustering algorithm with a linear complexity are even more promising than the ones from sofisticated K-means algorithms, which are running in polynomial times. But the processing of the data of our size would not even be possible by these more complex algorithms, if we didn't divide the work so efficiently. That means we would not have a comparison for the strategies and could not state that our linear complexity clustering algorithm provides very useful results, even if compared to these more complex algorithms.

There is another huge benefit of our thesis. We effectively created a pattern that makes clustering or any other processing of spherical data very efficient even for high complexity algorithms as we are able to slice the data space into chunks, that can be processed separately, without degrading the result quality.

\bibliographystyle{iso690}
\bibliography{Diplomka}

\setsecnumdepth{all}
\appendix

\chapter{Acronyms}
\begin{description}
	\item[BAT] Binary Association Table
	\item[DaCHS] Data Center Helper Suite
	\item[EM] Expectation maximization
	\item[GAVO] German Astrophysical Virtual Observatory
	\item[GPU] Graphical Processing Unit
	\item[HEALPix] Hierarchical Equal Area isoLatitude Pixelization of a sphere
	\item[HTM] Hierarchical Triangular Mesh
	\item[IVOA] International Virtual Observatory Alliance
	\item[LSST] Large Synoptic Survey Telescope
	\item[MPI] Message Passing Interface
	\item[OSPS] Ondřejov Southern Sky Photometry Survey
	\item[Q3C] Quad Tree Cube
	\item[SMC] Small Magellanic Cloud
	\item[PL/SQL] Procedural Language - Structured Query Language
	\item[PPMXL] Catalog of positions and proper motions on the ICRS
	\item[UCAC4] The Fourth US Naval Observatory CCD Astrograph Catalog
	\item[UDF] User Defined Function
	\item[VO] Virtual Observatory
\end{description}

\chapter{Contents of enclosed CD}


\begin{figure}
	\dirtree{%
		.1 readme.txt\DTcomment{the file with CD contents description}.
		.1 src\DTcomment{the directory of source codes}.
		.2 src\DTcomment{implementation sources}.
		.2 thesis\DTcomment{the directory of \LaTeX{} source codes of the thesis}.
		.1 text\DTcomment{the thesis text directory}.
		.2 thesis.pdf\DTcomment{the thesis text in PDF format}.
	}
\end{figure}

\chapter {Source codes}
\section {buildChunksFromCoordinates}
\label{codeChunks}
\begin{lstlisting}
int ChunkOperator::buildChunksFromCoordinates(int64 base1Nside, int64 base2Nside) {

    // the algorithm needs a vector of pointings to work, so we need to do some conversion
    vector<pointing> observations;
    for (size_t i = 0; i < obsCoords->size(); i++) {
        observations.push_back(pointing());
        observations[i].theta = (90 - (*obsCoords)[i].dec) * M_PI / 180; // colatitude in radian, some function of observations_degrees[i]
        observations[i].phi = (*obsCoords)[i].ra * M_PI / 180; // longitude in radian, some function of observations_degrees[i]
    }
    size_t noObservations = obsCoords->size();
    HEALPix_Base2 base1(base1Nside, NEST, SET_NSIDE);
    HEALPix_Base2 base2(base2Nside, NEST, SET_NSIDE);

    for (size_t i = 0; i < noObservations; i++) {
        // first see into which job the observation falls
        int64 idx_lores = base1.ang2pix(observations[i]);
        (*obsInCell)[idx_lores].push_back(&(*obsCoords)[i]);
        coords_by_pix_it got = obsInOverlap->find(idx_lores);
        if (got == obsInOverlap->end()){
            (*obsInOverlap)[idx_lores] = vector<Coordinate *> ();
        }

        // now check whether the surroundings of the observation touch neighbouring jobs
        int64 idx_hires = base2.ang2pix(observations[i]);

        fix_arr<int64, 8> neighbors;
        base2.neighbors(idx_hires, neighbors);
        for (size_t j = 0; j < 8; ++j) {
            if (neighbors[j] >= 0) {
                int64 nbidx_lores = base1.ang2pix(base2.pix2ang(neighbors[j]));

                if (nbidx_lores != idx_lores) { // touches a neighbour cell  
                    if ((*obsInOverlap)[nbidx_lores].empty() ||
                            (*obsInOverlap)[nbidx_lores].back() != &(*obsCoords)[i]) {
                        (*obsInOverlap)[nbidx_lores].push_back(&(*obsCoords)[i]);
                    }
                }
            }
        }
    }
    return 0;
}
\end{lstlisting}

\end{document}